\documentclass[fleqn,usenatbib]{mnras}

\usepackage{newtxtext,newtxmath}

\usepackage[T1]{fontenc}
\usepackage{xcolor}

\DeclareRobustCommand{\VAN}[3]{#2}
\let\VANthebibliography\thebibliography
\def\thebibliography{\DeclareRobustCommand{\VAN}[3]{##3}\VANthebibliography}

\usepackage{graphicx}	
\usepackage{amsmath}	
\usepackage{color}
\usepackage{times}
\usepackage{hyperref}                  
\hypersetup{colorlinks=true, urlcolor=blue, linkcolor=blue,  citecolor=blue}
\usepackage{bm}
\usepackage{multirow}
\usepackage{enumitem}

\title[SNe~Ia ejecta velocities]{Supernovae Ia ejecta velocities and host galaxy environments:\\
                                 the role of survey-selection effects}

\author[L.~V.~Barkhudaryan~et~al.]{L.~V.~Barkhudaryan,$^{1}$
A.~A.~Hakobyan,$^{1}$\thanks{E-mail: \href{mailto:artur.hakobyan@yerphi.am}{artur.hakobyan@yerphi.am}}
M.~H.~Gevorgyan,$^{1}$
D.~Kunth,$^{2}$
G.~A.~Mamon,$^{2}$
\newauthor
U.~Burgaz,$^{3}$
V.~Adibekyan,$^{4,5}$
Zh.~R.~Martirosyan$^{6}$
and A.~G.~Karapetyan$^{1}$
\\
$^{1}$Center for Cosmology and Astrophysics, Alikhanian National Science Laboratory, 2 Alikhanian Brothers Str, 0036 Yerevan, Armenia\\
$^{2}$Institut d'Astrophysique de Paris (UMR 7095: CNRS and Sorbonne Universit\'{e}), 98 bis bd Arago, F-75014 Paris, France\\
$^{3}$School of Physics, Trinity College Dublin, College Green, Dublin 2, Ireland\\
$^{4}$Instituto de Astrof\'isica e Ci\^encias do Espa\c{c}o, Universidade do Porto, CAUP, Rua das Estrelas, 4150-762 Porto, Portugal\\
$^{5}$Departamento de F\'{\i}sica e Astronomia, Faculdade de Ci\^encias, Universidade do Porto, Rua do Campo  Alegre, 4169-007 Porto, Portugal\\
$^{6}$Institute of Physics, Yerevan State University, 1 Alex Manoogian Str., 0025 Yerevan, Armenia
}

\date{Accepted - - -. Received - - -; in original form - - -}

\pubyear{\the\year{}}

\begin{document}
\label{firstpage}
\pagerange{\pageref{firstpage}--\pageref{lastpage}}
\maketitle

\begin{abstract}

The origin of the near-maximum-light \mbox{Si\,{\sc ii}}~$\lambda$6355 velocity diversity
among supernovae Ia (SNe~Ia) remains uncertain.
Previous studies have suggested that high-velocity (HV) and normal-velocity (NV) SNe~Ia
occupy systematically different host galaxy environments, implying differences in their progenitor populations.
We re-examine this hypothesis using a sample of 354 nearby ($z\leq0.04$)
spectroscopically normal SNe~Ia, comprising 239 NV and 115 HV events, identified in targeted and untargeted surveys.
For each SN, we determine the host galaxy morphology, galactocentric distance, and physical size,
and obtain homogeneous stellar mass estimates while assessing the influence of survey-selection effects.
The \mbox{Si\,{\sc ii}} velocity distribution is well described by two Gaussian components,
confirming the NV and HV populations.
We find no statistically significant differences between the galactocentric distance,
host galaxy size, or stellar mass distributions of the two velocity subgroups.
The environmental trends reported in previous studies are recovered only with marginal statistical significance
in the targeted subsample, whereas they disappear almost entirely in the untargeted subsample,
consistent with survey-selection effects playing a major role in the previously inferred environmental
differences between the NV and HV events.
Our results weaken the interpretation that the observed \mbox{Si\,{\sc ii}} velocity diversity is primarily driven by systematic differences in the global host properties examined here. Instead, they favour a scenario in which intrinsic explosion asymmetries and viewing-angle effects account for a substantial fraction of the observed velocity diversity, without excluding a possible contribution from local environmental conditions or progenitor properties.

\end{abstract}




\section{Introduction}
\label{intro}
\defcitealias{2013Sci...340..170W}{W13}
\defcitealias{2020ApJ...895L...5P}{P20}

Type~Ia supernovae (SNe~Ia) are widely believed to result from the thermonuclear disruption of
carbon--oxygen white dwarfs in close binary systems, although their progenitor systems and explosion
mechanisms remain poorly understood \citep[e.g.][]{2012PASA...29..447M,2014ARA&A..52..107M,2025A&ARv..33....1R}.
While their remarkable luminosity makes them powerful cosmological distance indicators
\citep[e.g.][]{1998AJ....116.1009R,1999ApJ...517..565P},
the diversity of their photometric and spectroscopic properties
\citep[e.g.][]{2017hsn..book..317T,2025A&ARv..33....1R}
indicates that multiple progenitor channels and/or explosion mechanisms may contribute to the SN~Ia population
\citep[e.g.][]{2000ARA&A..38..191H,2018PhR...736....1L,2023RAA....23h2001L},
motivating efforts to identify the physical origin of this diversity.

Among the numerous observables characterising SNe~Ia, the photospheric velocity inferred from the
\mbox{Si\,{\sc ii}}~$\lambda$6355 absorption feature near maximum light has emerged as one of the
most important spectroscopic diagnostics of the explosion.
It traces the expansion velocity of the line-forming layers in the ejecta and is sensitive to the explosion energetics,
density structure, chemical composition, and viewing geometry
\citep[e.g.][]{2005ApJ...623.1011B,2007Sci...315..825M,2010Natur.466...82M,2013MNRAS.429.1156S}.
Early studies showed that the distribution of near-maximum \mbox{Si\,{\sc ii}} velocities
is bimodal and can be broadly divided into normal-velocity (NV) and high-velocity (HV)
populations using a threshold of $\sim12000$ km~s$^{-1}$
(\citealt{2009ApJ...699L.139W,2013Sci...340..170W}, hereafter \citetalias{2013Sci...340..170W}).
The physical origin of this velocity diversity, however, remains uncertain.
Several studies have suggested that the observed NV--HV diversity may be connected to progenitor properties. \citetalias{2013Sci...340..170W} interpreted the central concentration and host galaxy differences of HV SNe~Ia as evidence for systematically younger and more metal-rich progenitor systems, whereas \citet{2015MNRAS.446..354P} and \citet[][hereafter \citetalias{2020ApJ...895L...5P}]{2020ApJ...895L...5P}
reported a preference for more massive hosts and discussed metallicity as a possible underlying factor.
Other authors have argued that a substantial fraction of the observed diversity can instead be explained by intrinsic explosion asymmetries
and viewing-angle effects without invoking distinct progenitor populations
\citep[e.g.][]{2010Natur.466...82M,2020MNRAS.499.5325Z}.

Host galaxy environments provide an important observational avenue for investigating the origin of the
\mbox{Si\,{\sc ii}} velocity diversity. Global galaxy properties, such as morphology, stellar mass,
luminosity, and star-formation activity, are statistically linked to the ages, metallicities,
and evolutionary histories of their stellar populations
\citep[e.g.][]{2015A&A...581A.103G},
while the galactocentric positions of SNe probe local variations in these quantities through
the well-established radial gradients observed in galaxies
\citep[e.g.][]{2014A&A...572A..38G,2018ApJ...855..107G,2015MNRAS.448..732A,2016MNRAS.456.2848H,2021MNRAS.505L..52H}.
Consequently, comparisons of the environments of NV and HV SNe~Ia offer an indirect means of testing
whether the observed ejecta velocity diversity is primarily driven by differences in progenitor populations
or instead reflects intrinsic properties of the explosion itself.

One of the first comprehensive investigations of the relation between \mbox{Si\,{\sc ii}} velocities
and host galaxy environments was carried out by \citetalias{2013Sci...340..170W},
who analysed a sample of 188 nearby ($z<0.05$; mean $z=0.023$) SNe~Ia from the
Lick Observatory Supernova Search (LOSS), including 123 spectroscopically normal events
(83 NV and 40 HV SNe~Ia). The photospheric velocities at maximum light were derived from
\mbox{Si\,{\sc ii}}~$\lambda$6355 measurements obtained within $-7\leq t\leq+7$ d of maximum
using template-based phase corrections.
Galactocentric distances of 120 SNe~Ia were quantified by the normalised galactocentric radius,
$R_{\rm SN}/R_{25}$, where $R_{25}$ is the semimajor axis of the $B$-band 25 mag arcsec$^{-2}$ isophote.
SN deprojected distances were adopted for moderately inclined disc galaxies
$(i<70^{\circ})$, whereas projected distances were used for elliptical and highly inclined disc galaxies.
They found that HV SNe~Ia are significantly more centrally concentrated than NV events
and preferentially occur in larger and more luminous host galaxies
(in all cases $P_{\rm KS}=0.005$).
These findings were interpreted as evidence that HV SNe~Ia arise from systematically younger
and more metal-rich progenitor populations, suggesting that the observed ejecta velocity diversity
reflects intrinsic differences in the progenitor systems rather than being produced solely by
explosion asymmetries or viewing-angle effects
\citep[e.g.][]{2010Natur.466...82M,2018MNRAS.477.3567M,2020MNRAS.499.5325Z}.

Subsequently, \citet{2015MNRAS.446..354P} investigated the relation between
\mbox{Si\,{\sc ii}} velocities and host galaxy properties for 122 SNe~Ia
($z<0.09$) discovered by the Palomar Transient Factory (PTF).
Using spectra obtained within $\pm5$ d of maximum light, they examined the projected
galactocentric distances of 98 SNe~Ia, normalised to the galaxy radius enclosing
90 per cent of the total light $(R_{\rm SN}/R_{90})$, together with stellar masses
available for 102 hosts.
In contrast to \citetalias{2013Sci...340..170W}, they found no statistically significant
difference between the radial distributions of NV and HV SNe~Ia, although HV events
were preferentially associated with more massive hosts $(M_\star > 3 \times 10^{9}~{\rm M_{\odot}})$.

Using an enlarged sample of 217 SNe~Ia with host galaxy stellar mass estimates,
extending to $z<0.2$, \citetalias{2020ApJ...895L...5P} revisited the connection between
\mbox{Si\,{\sc ii}} velocity and host galaxy properties.
Their sample combined SNe~Ia discovered by the PTF with additional objects
from the Berkeley SN~Ia Program \citep{2012MNRAS.425.1819S,2015MNRAS.451.1973S}.
Restricting the analysis to spectra obtained within $\pm3$ d of maximum light,
they found that HV SNe~Ia occur preferentially in more massive host galaxies,
thereby reinforcing the proposed environmental distinction between HV and NV populations
\citep[see also][]{2021ApJ...923..267D},
and suggested that progenitor metallicity is likely the dominant factor governing the
observed velocity diversity.

More recently, extending the environmental context,
\citet{2023arXiv230410601N} analysed a sample of 74 nearby ($z<0.04$) SNe~Ia
compiled from several galaxy-targeted surveys, including LOSS, the Harvard--Smithsonian Center for Astrophysics
Supernova Program \citep{2009ApJ...700..331H}, and the Carnegie Supernova Project \citep{2010AJ....139..519C}.
Projected galactocentric distances for 69 SNe~Ia were normalised to the host galaxy size,
$R_{\rm SN}/R_{\rm gal}$, where $R_{\rm gal}$ is the semimajor axis of the
Two Micron All-Sky Survey (2MASS) $K$-band 20 mag~arcsec$^{-2}$ isophote.
They found that HV SNe~Ia are significantly more centrally concentrated than NV events
($P_{\rm AD}=0.003$), consistent with the results of \citetalias{2013Sci...340..170W}.
However, no statistically significant differences were detected in global host properties,
including stellar mass and stellar population age,
suggesting that the two velocity subgroups inhabit broadly similar global environments.

\citet{2025A&A...698A.305M} compiled a complete sample of nearby SNe within 40~Mpc
from modern wide-field untargeted surveys, including 69 spectroscopically classified SNe~Ia.
Among the spectroscopically normal events, 13 were classified as HV and 36 as NV
(HV:NV ratio $=1:2.77$, consistent with earlier estimates; e.g. \citealt{2009ApJ...699L.139W}).
Radial distributions were analysed using projected and $R_{25}$-normalised galactocentric distances for
SNe in elliptical and lenticular hosts, and deprojected and $R_{25}$-normalised distances for SNe in spiral galaxies.
They found only a tendency for HV SNe~Ia to be more centrally concentrated than NV events,
which did not reach statistical significance.
Similarly, no significant difference was found between the host stellar masses of the two velocity subgroups.

Finally, with the advent of wide-field untargeted surveys, substantially larger and homogeneous samples have become available.
Using 293 spectroscopically normal SNe~Ia with $z < 0.06$ from the ZTF DR2, \citet{2025A&A...694A..13B} measured \mbox{Si\,{\sc ii}} velocities
within the phase interval $-5 \leq t \leq +5$~d.
Addressing the influence of host contamination on velocity measurements,
they examined the relation between the \mbox{Si\,{\sc ii}}~$\lambda$6355 velocity and $d_{\rm DLR}$,
defined as the SN offset scaled by the Directional Light Radius (DLR), i.e. the elliptical radius of host galaxy
measured in the direction of the SN position \citep[see, e.g.][]{2016AJ....152..154G}.
No statistically significant difference was found between the radial distributions of HV and NV events $(P_{\rm KS} = 0.85)$.
Furthermore, their results indicate that HV SNe~Ia are not confined to massive galaxies, but also occur in hosts of
intermediate and relatively low stellar mass. They argued that the lack of HV events in earlier studies is
thus most likely attributable to limited sample sizes and selection biases favouring discoveries in more massive systems,
rather than an intrinsic absence of HV SNe~Ia from low-mass hosts.

Thus, despite more than a decade of observational studies, the environmental origin of the
NV--HV dichotomy, first reported by \citetalias{2013Sci...340..170W}, remains unsettled.
Previous investigations have employed heterogeneous SN samples, different host galaxy measurements,
and various definitions of galactocentric distance, while the increasing contribution of untargeted
SN surveys has revealed substantially weaker environmental differences than initially reported.
These developments raise the question of whether the published discrepancies primarily reflect intrinsic
differences between the NV and HV events or are, at least in part, the consequence of observational
selection effects and methodological differences.

In this paper, we investigate the environmental properties of 354 nearby ($z\leq0.04$)
spectroscopically normal SNe~Ia drawn from both targeted and untargeted SN surveys.
The increased sample size enables statistically robust comparisons between the NV and HV subgroups
across a broad range of host galaxy environments, including different morphological types,
galactocentric distances, physical sizes, and stellar masses.
We derive the host galaxy parameters using homogeneous measurements and a well-tested methodology
developed in our previous studies
\citep{2009A&A...508.1259H,2012A&A...544A..81H,2016MNRAS.456.2848H,2017MNRAS.471.1390H,
2019MNRAS.490..718B,2020MNRAS.499.1424H,2021MNRAS.505L..52H,2023MNRAS.520L..21B},
thereby minimising methodological inconsistencies among the different environmental indicators.
Our primary goals are to reassess the reported environmental differences between NV and HV SNe~Ia
using homogeneous measurements for a nearby SN~Ia sample and to quantify the extent to which
survey-selection effects influence the observed \mbox{Si\,{\sc ii}} velocity distribution.

The remainder of this paper is organised as follows.
Sect.~\ref{samplered} describes the SN~Ia sample, the host galaxy data,
and the procedures adopted to derive the environmental parameters.
Sect.~\ref{RESults} presents the results and discusses their implications in the context of previous studies.
Our main conclusions are summarised in Sect.~\ref{Concl}.
Throughout this paper, we adopt a cosmological model with
$\Omega_{\rm m}=0.27$, $\Omega_{\Lambda}=0.73$, and
$H_0=73$ km~s$^{-1}$~Mpc$^{-1}$.

\section{Sample selection and reduction}
\label{samplered}

\subsection{SN~Ia data}
\label{Iadata}

To construct a statistically robust sample and ensure reliable measurements of galactocentric distances
for spectroscopically normal SNe~Ia, we compiled a dataset of relatively nearby events with redshift
$z \leq 0.04$ and published expansion velocity measurements of the \mbox{Si\,{\sc ii}}~$\lambda$6355 absorption feature.
The low-redshift cut enables accurate SN-host positional analyses.
A quantitative illustration of the dependence of the angular host galaxy radii on redshift is provided in Appendix~\ref{app:angular-size}.\footnote{The adopted $z \leq 0.04$ limit is a practical selection criterion based on host galaxy characterisation requirements.
In the highest-redshift bin, the mean angular galaxy radius is approximately 17 arcsec, above the 15 arcsec practical limit indicated by our previous experience.}
We collected \mbox{Si\,{\sc ii}} velocity measurements from the literature, including
\citet{2012MNRAS.425.1819S,2012AJ....143..126B}; \citetalias{2013Sci...340..170W};
\citet{2013ApJ...773...53F,2015ApJS..220...20Z,2020ApJ...901..154B,2021ApJ...923..267D,2024ApJ...967...20M,2025A&A...694A...9B}.
Each SN is required to have at least one \mbox{Si\,{\sc ii}}~$\lambda$6355 velocity measurement
within a phase interval of $-5 \leq t \leq +5$~d relative to $B$-band maximum light.
For SNe~Ia with multiple measurements within the phase range, we retained the velocity measured closest to maximum light.
The compiled sample comprises 404 SNe~Ia with reliable \mbox{Si\,{\sc ii}} velocity measurements.

To ensure homogeneity, we convert all measurements to velocities at maximum light,
using the empirical velocity-phase relation introduced by \citet{2011ApJ...742...89F}.
This approach mitigates phase-dependent systematics and allows direct comparison of ejecta velocities across the sample.
An object-by-object comparison between the original velocity measurements and the phase-corrected velocities is presented in Appendix~\ref{app:phase-correction}.
For the adopted phase-restricted sample, the corrections are modest and show no systematic dependence on the available SN~Ia light-curve decline-rate groups.

All 404 SNe~Ia were required to have a published spectroscopic subclassification confirming them as normal SNe~Ia.
To compile this information, we performed a comprehensive literature search following
the approach outlined in \citet{2020MNRAS.499.1424H}.
As the primary source of spectroscopic data and references,
we used the Weizmann Interactive Supernova Data Repository\footnote{\href{http://www.wiserep.org/}{http://www.wiserep.org/}}
\citep[WISeREP;][]{2012PASP..124..668Y},
an extensive and interactive archive of SN spectra along with associated references.
We further cross-checked classifications using reports from
the Central Bureau for Astronomical Telegrams\footnote{\href{http://www.cbat.eps.harvard.edu/iau/cbat.html}{http://www.cbat.eps.harvard.edu/iau/cbat.html}}
(CBAT)
and the Astronomer's Telegram\footnote{\href{http://www.astronomerstelegram.org/}{http://www.astronomerstelegram.org/}}
(ATEL), as well as additional literature sources
\citep[e.g.][]{2012MNRAS.425.1819S,2013ApJ...773...53F}.
The use of multiple sources allowed us to ensure a high degree of accuracy in the subclassification of these SNe~Ia.

To robustly identify and characterize the host galaxies of the 404 SNe~Ia across a broad range of environments,
we cross-matched the SN coordinates with imaging data from the nineteenth data release of the Sloan Digital Sky Survey
\citep[DR19;][]{2026ApJS..285....9S}, the fourth data release of SkyMapper \citep[DR4;][]{2024PASA...41...61O},
and the second data release of Pan-STARRS \citep[DR2;][]{2020ApJS..251....7F}.
The combined sky coverage of these surveys is effectively all-sky, enabling host identification for the full SN sample.

For nine SNe (e.g. SN~2000dk, SN~2006bw, and SN~2008bf), the nearly equal projected separations from
two or more nearby elliptical or lenticular galaxies, or their location in close projection between potential hosts,
prevented an unambiguous host galaxy identification.
These objects were therefore excluded from the sample to avoid introducing systematic uncertainties into the environmental analysis.

To assess the presence of strong morphological disturbances among the remaining host galaxies,
we visually inspected colour-composite images from the SDSS, Pan-STARRS, and SkyMapper surveys.
The composite images were constructed using the $g$-, $r$-, and $i$-band filters available in each survey.
Forty-one host galaxies (e.g. UGC~3781, UGC~9425, and NGC~7469) exhibit clear signatures of strong disturbance,
such as prominent tidal interactions, ongoing mergers, or well-defined post-merger features,
according to the criteria of \citet{2014MNRAS.444.2428H}.
These systems were excluded from further analysis, as our study focuses specifically on
the distribution of SNe~Ia in morphologically undisturbed host galaxies.

\begin{figure}
\begin{center}$
\begin{array}{@{\hspace{0mm}}c@{\hspace{0mm}}}
\includegraphics[width=1\hsize]{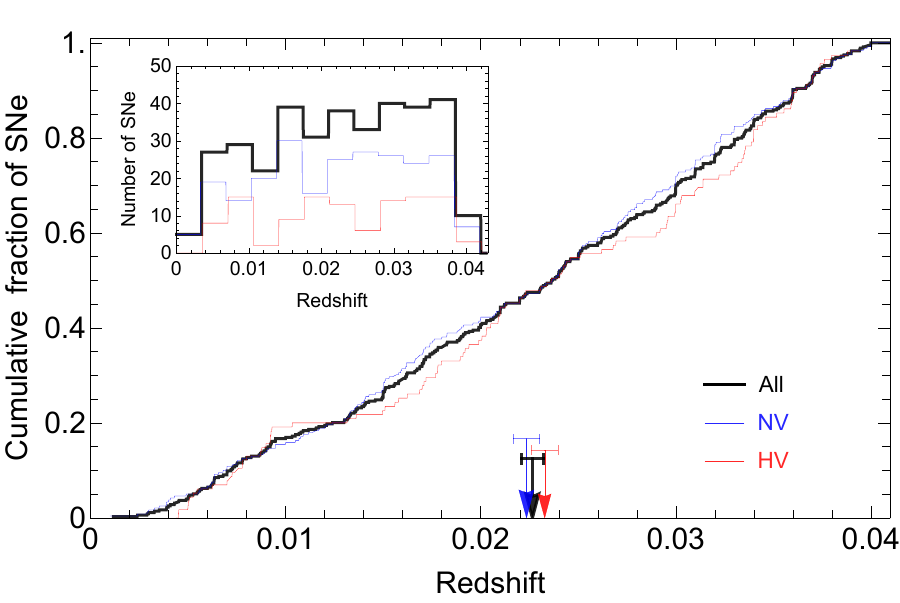}
\end{array}$
\end{center}
\caption{Cumulative redshift distribution of spectroscopically normal (NV and HV) SNe~Ia.
         The inset shows the corresponding redshift histograms.
         Arrows indicate the mean redshift of each subgroup,
         with error bars representing the standard error of the mean.}
\label{Redshift_valid}
\end{figure}

After applying the above selection criteria, the final sample comprises 354 spectroscopically normal SNe~Ia.
Among these, 239 are classified as NV ($V_{\rm Si} < 12000$ km~s$^{-1}$) and
115 as HV ($V_{\rm Si} \geq 12000$ km~s$^{-1}$).
The cumulative redshift distributions and corresponding histograms of the NV and HV subgroups are presented in Fig.~\ref{Redshift_valid}.
Statistical differences between the redshift distributions of these two subgroups were evaluated using
the two-sample Kolmogorov--Smirnov (KS) and Anderson--Darling (AD) tests.\footnote{The KS and AD tests
are non-parametric methods designed to evaluate whether two samples are
consistent with being drawn from the same parent distribution.
The KS test quantifies the maximum absolute difference between the empirical
cumulative distribution functions (CDFs) of the samples,
whereas the AD test places greater weight on discrepancies in the tails of the distributions.
Both tests yield a $P$-value representing the probability of obtaining a difference at least as extreme as
the observed one under the null hypothesis that the samples originate from the same distribution.
Smaller $P$-values therefore indicate stronger evidence against the null hypothesis.
Throughout this study, we adopt a significance threshold of 5 per cent $(P<0.05)$.}
The results show no statistically significant difference between the redshift distributions of NV and HV SNe~Ia.
Both tests yield $P$-values greater than 0.5, well above the adopted significance threshold $(P<0.05)$.
The results provide no statistically significant evidence that the redshift distributions of the NV and HV subgroups differ across the adopted redshift range,
and therefore show no indication of a redshift-dependent imbalance between the two subgroups within our sample.

As a consistency check on the relative NV/HV fractions in our sample,
we compared the fraction of HV events with that reported for a volume-limited complete sample within 40~Mpc,
discovered primarily by modern wide-field untargeted surveys between 2016 and 2023 \citep{2025A&A...698A.305M}.
In our sample, $32\pm3$ per cent of SNe~Ia are classified as HV events, which is fully consistent,
within uncertainties, with the HV fraction of $27^{+8}_{-7}$ per cent reported by \citet{2025A&A...698A.305M}.
This close agreement with the volume-limited complete sample indicates that our dataset does not exhibit significant
artificial deficits or excesses of NV or HV SNe~Ia.

In recent years, the importance of using untargeted SN samples in statistical analyses has become
increasingly evident \citep[e.g.][]{2017PASP..129j4502K,2025A&A...694A...1R,2025A&A...694A...9B}.
Early SN surveys, limited by the relatively small fields of view of available telescopes,
preferentially targeted massive, luminous galaxies -- often characterised by high star-formation rates --
in order to maximize detection efficiency.
Because massive galaxies typically possess higher gas-phase metallicities \citep[e.g.][]{2004ApJ...613..898T},
such strategies introduced systematic environmental biases into the resulting SN host-galaxy samples
\citep[e.g.][]{2016A&A...591A..48G,2021ApJ...923..267D}.

The advent of wide-field time-domain facilities has fundamentally transformed SN discovery.
Modern surveys such as the All-Sky Automated Survey for Supernovae \citep[ASAS-SN;][]{2017PASP..129j4502K} and
the Zwicky Transient Facility \citep[ZTF;][]{2025A&A...694A...1R,2025A&A...694A..10D} repeatedly scan large fractions of
the sky without preselecting host galaxies.
Consequently, SNe discovered in these untargeted or quasi-untargeted surveys are substantially less affected by host galaxy selection effects.

In this context, we examined the discovery information for the 354 SNe~Ia in our sample using multiple sources,
including the Asiago Supernova Catalogue \citep[ASC;][]{1999A&AS..139..531B}, ZTF, ASAS-SN, WISeREP,
and additional survey databases and literature where available.
Each SN~Ia was classified according to its discovery strategy, distinguishing between targeted surveys,
which monitor predefined galaxy samples, and untargeted surveys, which employ wide-field search strategies with minimal host preselection.
Based on this classification, 211 SNe~Ia were identified as originating from targeted surveys, while 143 were discovered by untargeted surveys.

\subsection{Host galaxies}
\label{Hostdata}

We identified 351 distinct host galaxies: 348 galaxies host a single SN~Ia, whereas three galaxies host two events each.
Following the approach outlined by \citet{2012A&A...544A..81H},
we performed a detailed morphological classification of these 351 galaxies through
visual inspection of the colour-composite images described above.
The distribution of SNe~Ia across host morphological types is presented in Table~\ref{tabSNhostmorph}.

\begin{table}
  \centering
  \begin{minipage}{84mm}
  \caption{Distribution of spectroscopically normal (NV and HV) SNe~Ia
           across the morphological types of their host galaxies $(z \leq 0.04)$.}
  \tabcolsep 1.52pt
  \label{tabSNhostmorph}
  \begin{tabular}{lccccccccccccccc}
  \hline
  \multicolumn{1}{l}{Subgroup} &\multicolumn{1}{c}{E}&\multicolumn{1}{c}{E/S0}&\multicolumn{1}{c}{S0}&\multicolumn{1}{c}{S0/a}&\multicolumn{1}{c}{Sa}
  &\multicolumn{1}{c}{Sab}&\multicolumn{1}{c}{Sb}&\multicolumn{1}{c}{Sbc}&\multicolumn{1}{c}{Sc}&\multicolumn{1}{c}{Scd}
  &\multicolumn{1}{c}{Sd}&\multicolumn{1}{c}{Sdm}&\multicolumn{1}{c}{Sm}&\multicolumn{1}{c}{Im}&\multicolumn{1}{c}{All}\\
  \hline
  $N_{\rm NV}$ & 26 & 6 & 8 & 16 & 19 & 13 & 38 & 39 & 40 & 13 & 6 & 5 & 2 & 8 & 239 \\
  $N_{\rm HV}$ & 9 & 3 & 3 & 6 & 8 & 11 & 18 & 25 & 17 & 7 & 3 & 1 & 3 & 1 & 115 \\
  $N_{\rm All}$ & 35 & 9 & 11 & 22 & 27 & 24 & 56 & 64 & 57 & 20 & 9 & 6 & 5 & 9 & 354 \\
  \hline
  \end{tabular}
  \end{minipage}
\end{table}

To minimise the impact of projection effects and internal extinction, both of which are significant in
highly inclined galactic discs, and to ensure a reliable determination of the radial distributions of SNe~Ia
in S0--Sdm host galaxies, accurate inclination measurements and the application of appropriate inclination constraints
are essential \citep[e.g.][]{2016MNRAS.456.2848H}.
Following the methodology of \citet{2012A&A...544A..81H}, we constructed isophotal elliptical apertures at
the 25 mag arcsec$^{-2}$ surface brightness level on the $g$-band images of each host galaxy.
These apertures were used to measure the semimajor $(R_{25})$ and semiminor $(Z_{25})$ axes.
Inclinations were calculated from the axis ratio $(R_{25}/Z_{25})$ and morphological type
using the approach of \citet{1997A&AS..124..109P}.
Within our sample, 236 SNe~Ia are hosted by S0--Sdm galaxies with $i < 70^\circ$,
while 60 reside in more highly inclined systems.

For all 354 SNe~Ia, the projected offsets from the host galaxy nuclei were calculated exclusively from the
precise equatorial coordinates of the SN and the nucleus (photometric centre) of its host galaxy. The offsets reported in
the ASC, which are generally more reliable than those compiled from heterogeneous sources \citep{2012A&A...544A..81H},
were used only as a consistency check.

In elliptical galaxies, the true three-dimensional galactocentric distance of a SN cannot be directly determined
from its observed angular offsets in right ascension $(\Delta \alpha)$ and declination $(\Delta \delta)$.
We therefore computed the projected distance in the plane of the sky as
$R_{\rm SN} = \sqrt{(\Delta \alpha)^2 + (\Delta \delta)^2}$,
which provides a lower limit to the intrinsic galactocentric distance.
These projected distances were then normalised by the $g$-band isophotal radii of the host galaxies, $R_{25}$,
yielding the dimensionless relative galactocentric distances $R_{\rm SN}/R_{25}$ \citep[e.g.][]{2019MNRAS.490..718B}.
Note that the $R_{25}$ radii were corrected for Galactic foreground extinction \citep{2011ApJ...737..103S}
and for internal extinction within the hosts \citep{1995A&A...296...64B}.
The $R_{25}$ values in kpc were derived from the angular $R_{25}$ measurements using galaxy distances estimated from
recession velocities corrected both to the centroid of the Local Group \citep{1977ApJ...217..903Y} and for
the infall of the Local Group towards the Virgo Cluster \citep{1998A&A...340...21T,2002A&A...393...57T}.

The above procedure can, in principle, be applied to the galactocentric distances of SNe~Ia in host galaxies of all morphological types.
However, more geometrically appropriate approaches are available for specific cases,
particularly when the spatial distribution of the underlying stellar populations is taken into account.
For example, in spiral galaxies the majority of SNe~Ia are observed to occur in disc components rather than in bulges
\citep[e.g.][]{1997ApJ...483L..29W,2015MNRAS.448..732A,2016MNRAS.456.2848H,2021MNRAS.505L..52H}.
Accordingly, we applied inclination corrections to the observed galactocentric distances of SNe~Ia,
following the procedure described in \citet{2009A&A...508.1259H}, and subsequently normalised the deprojected $R_{\rm SN}$ radii by
the corresponding host $R_{25}$ values.
This deprojection assumes that SNe are confined to infinitely thin discs in S0/a--Sdm galaxies.
However, for highly inclined systems, particularly those approaching edge-on orientations,
such a correction becomes unreliable due to projection effects associated with the finite vertical thickness of real galactic discs
\citep[see, e.g.][]{2017MNRAS.471.1390H,2023MNRAS.520L..21B}.

The same inclination correction, subject to the limitations discussed above, can also be applied to SNe~Ia
hosted by lenticular (S0) galaxies. However, a non-negligible fraction of SNe~Ia in these systems may originate from
extended bulge components, potentially biasing the inferred radial distributions \citep[e.g.][]{2016MNRAS.456.2848H}.

For SNe~Ia hosted by Sm--Im galaxies, the projected galactocentric distances were computed from the measured offsets
relative to the photometric centres (i.e. the brightest regions) of the hosts and normalised by the corresponding $R_{25}$ values.

To obtain homogeneous stellar mass estimates for the 354 SN~Ia host galaxies in our sample,
we cross-matched them with the Revised Galaxy List for the Advanced Detector Era \citep[REGALADE;][]{2026A&A...706A.284T}.
REGALADE is a high-purity, high-completeness all-sky galaxy catalogue specifically developed for transient and multi-messenger astrophysics,
combining major galaxy catalogues, deep imaging surveys, and spectroscopic, photometric, and redshift-independent distance measurements into a unified database.
It provides homogeneous estimates of galaxy stellar masses for the vast majority of its nearly 80 million galaxies,
making it particularly well suited for statistical studies of SN host galaxy properties.

REGALADE contains stellar mass estimates for 342 of the 354 SN~Ia host galaxies in our sample ($\sim97$ per cent),
which we cross-checked against the corresponding survey imaging data to ensure the reliability of the host identifications.
For the remaining 12 host galaxies, we derived stellar masses using spectral energy distribution (SED) fitting based
on multiwavelength photometry spanning the ultraviolet (UV) to the near-infrared (NIR).
UV magnitudes were obtained from the Galaxy Evolution Explorer \citep[GALEX;][]{2005ApJ...619L...1M};
optical $u$, $g$, $r$, $i$, and $z$ photometry was compiled from SDSS, SkyMapper, and Pan-STARRS;
and NIR $J$, $H$, and $K_{\rm s}$ measurements were retrieved from the 2MASS \citep[][]{2006AJ....131.1163S}.
The photometric data were primarily obtained from the NASA/IPAC Extragalactic
Database\footnote{\href{https://ned.ipac.caltech.edu/}{https://ned.ipac.caltech.edu/}} (NED) and supplemented,
where available, with measurements from \citet{2026ApJ...997...38N}.
All magnitudes were corrected for Galactic foreground extinction using the extinction maps of \citet{2011ApJ...737..103S}.

Using the spectroscopic redshift of each of the remaining 12 host galaxies together with the multi-band photometry and
morphological information, we estimated stellar masses with the \textsc{pegase.2} evolutionary synthesis code
\citep{1997A&A...326..950F,1999astro.ph.12179F}.
The observed galaxy colours were fitted with synthetic SEDs corresponding to nine spectral-morphological types
(SB, Im, Sd, Sc, Sbc, Sb, Sa, S0, and E), representing different star-formation histories and stellar populations.
A Salpeter initial mass function (IMF) was adopted throughout the fitting procedure.
As with any SED-fitting technique, the derived stellar masses depend on assumptions regarding the adopted IMF,
star-formation history, and dust attenuation, resulting in typical systematic uncertainties of $\sim0.1$--$0.3$ dex.
A detailed description of the fitting procedure, including the filter-convolution and template-smoothing algorithms,
is available in the \textsc{pegase.2} documentation.\footnote{\href{https://www2.iap.fr/pegase/}{https://www2.iap.fr/pegase/}}

\section{Results and discussion}
\label{RESults}

\subsection{Distribution of \mbox{Si\,{\sc ii}} velocities}
\label{RESults1}

The distribution of \mbox{Si\,{\sc ii}}~$\lambda$6355 velocities among SNe~Ia has been shown to
exhibit bimodality (e.g. \citetalias{2013Sci...340..170W}; \citealt{2020MNRAS.499.5325Z,2024MNRAS.532.1887P,2026MNRAS.546f2281H}).
In the analysis of \citetalias{2013Sci...340..170W}, the majority of 123 spectroscopically normal SNe~Ia were found to cluster within
$10000-12000$ ${\rm km~s^{-1}}$, with a less prominent HV tail extending to $\sim16000$ ${\rm km~s^{-1}}$.
\citetalias{2013Sci...340..170W} demonstrated that this distribution is well described by a bimodal Gaussian model comprising a narrow,
higher-amplitude component centred at 10800 ${\rm km~s^{-1}}$ (600 ${\rm km~s^{-1}}$ dispersion and $62\pm5$ per cent weight),
corresponding to the NV population, and a broader,
lower-amplitude component peaking at 13000 ${\rm km~s^{-1}}$ (1400 ${\rm km~s^{-1}}$ dispersion and $38\pm5$ per cent weight),
associated with HV events.
Adopting a threshold of 12000 ${\rm km~s^{-1}}$ to separate the two subgroups,
\citetalias{2013Sci...340..170W} classified 83 SNe~Ia as NV and 40 as HV,
corresponding to an HV fraction $(f_{\rm HV})$ of $32\pm4$ per cent among spectroscopically normal events.

More recently, analysing a combined untargeted sample of SNe~Ia limited to $z<0.09$
and a phase interval of $-5 \lesssim t \lesssim +5$~d,
\citet{2024MNRAS.532.1887P} reported mean \mbox{Si\,{\sc ii}} velocities of $10700\pm700$ ${\rm km~s^{-1}}$ for NV population and
$11900\pm1700$ ${\rm km~s^{-1}}$ for HV population, with corresponding Gaussian-component weights of $76\pm3$ and $24\pm3$ per cent, respectively.
Although the NV-component weight inferred in \citetalias{2013Sci...340..170W} is somewhat lower than that reported by
\citet{2024MNRAS.532.1887P}, this discrepancy may plausibly reflect differences in survey strategy and sample construction.
In particular, the targeted nature of the \citetalias{2013Sci...340..170W} sample, compared to the untargeted selection employed by
\citet{2024MNRAS.532.1887P}, may introduce systematic variations in the relative representation of the two velocity components.

\begin{figure}
\centering
    \includegraphics[width=1\hsize]{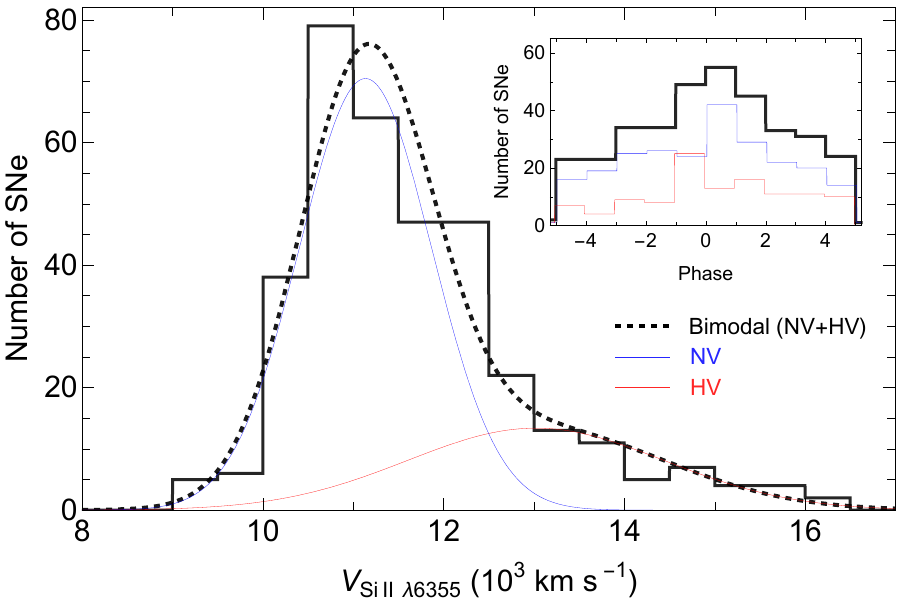}
    \caption{Histogram of \mbox{Si\,{\sc ii}} velocities measured near maximum light
    for our 354 spectroscopically normal SNe~Ia.
    The distribution is fitted with a bimodal Gaussian model:
    the blue and red curves denote the NV and HV components, respectively,
    while the dashed black curve represents the sum of the two Gaussians.
    The inset presents the phase distribution of the corresponding velocity measurements.}
    \label{Fits}
\end{figure}

In Fig.~\ref{Fits}, we present the distribution of \mbox{Si\,{\sc ii}} velocities for the
354 spectroscopically normal SNe~Ia in our sample.
The distribution is well described by a bimodal Gaussian model:
a narrower component centred at 11100 ${\rm km~s^{-1}}$ with 700 ${\rm km~s^{-1}}$ dispersion,
corresponding to the NV population, and a broader component peaking at 13000 ${\rm km~s^{-1}}$
with 1400 ${\rm km~s^{-1}}$ dispersion, associated with the HV population.\footnote{The bimodal fits
are performed using a maximum-likelihood estimation (MLE) technique,
following the formalism of \citet{2020MNRAS.499.5325Z}.
The optimisation routine was implemented in \textit{Wolfram Mathematica}.}
The NV and HV components account for $73^{+2}_{-3}$ and $27^{+3}_{-2}$ per cent of the bimodal
Gaussian-model weight, respectively, in excellent agreement with the corresponding component weights
reported by \citet{2024MNRAS.532.1887P} for their low-redshift ($z<0.09$) sample.
When our sample is restricted to the 211 SNe~Ia discovered by targeted surveys,
the NV and HV components account for $64^{+3}_{-4}$ and $36^{+4}_{-3}$ per cent of the bimodal Gaussian-model weight,
respectively, consistent with the values reported by \citetalias{2013Sci...340..170W}.
At this stage, we note that the nature of the sample selection (targeted versus untargeted)
appears to have a measurable impact on the inferred NV/HV component weights,
introducing systematic variations in the relative representation of the two velocity populations.
This possibility, together with other effects related to host galaxy properties,
will be examined in more detail in the following sections.

To quantify the effect of the individual velocity uncertainties on the bimodal fits,
we perform a Monte Carlo (MC) analysis in which the individual \mbox{Si\,{\sc ii}} velocities are perturbed
according to Gaussian distributions defined by the combined uncertainties,
$\sqrt{[Err(V_{\rm Si})]^{2}+220^{2}}$, where the $220$ ${\rm km~s^{-1}}$ phase-correction scatter reported by \citet{2011ApJ...742...89F} is added in quadrature to the reported velocity uncertainty.
For each of $10^4$ realisations,
we repeat the MLE fit and adopt the median and the 16th--84th percentile range as the final parameter estimates.
For the full sample, the median NV and HV component weights are $74\pm2$ and $26\pm2$ per cent, the
component means are $11100$ and $13000$ ${\rm km~s^{-1}}$, and the dispersions are $700$ and
$1400$ ${\rm km~s^{-1}}$. For the targeted subsample, the corresponding component weights are $66\pm3$ and
$34\pm3$ per cent, the component means are $11100$ and $12900$ ${\rm km~s^{-1}}$, and the dispersions are $700$ and $1400$ ${\rm km~s^{-1}}$.
Thus, the inferred bimodal parameters remain robust when the measurement uncertainties are propagated
and are consistent with the values reported by \citet{2024MNRAS.532.1887P} and \citetalias{2013Sci...340..170W}, respectively.
This also holds when the original, uncorrected velocities are used (see Appendix~\ref{app:uncorrected}).

\begin{table}
  \centering
  \begin{minipage}{84mm}
  \caption{Distribution of HV SN~Ia fractions ($f_{\rm HV}$, expressed in per cent),
           defined as the fraction of events with $V_{\rm Si} \geq 12000$ km~s$^{-1}$,
           across the different morphological ranges of the host galaxies in our sample.}
  \tabcolsep 11.2pt
  \label{tabSNhostmorphfHV}
  \begin{tabular}{lcccc}
  \hline
  \multicolumn{1}{l}{Un/Targeted} &\multicolumn{1}{c}{E--S0}&\multicolumn{1}{c}{S0/a--Sbc}&\multicolumn{1}{c}{Sc--Im}&\multicolumn{1}{c}{All}\\
  \hline
  All & \multicolumn{1}{c}{$27^{+7}_{-6}$} & \multicolumn{1}{c}{$35\pm4$} & \multicolumn{1}{c}{$30\pm5$} & \multicolumn{1}{c}{$32\pm3$} \\
  Targeted & \multicolumn{1}{c}{$24^{+12}_{-9}$} & \multicolumn{1}{c}{$36^{+5}_{-4}$} & \multicolumn{1}{c}{$34\pm4$} & \multicolumn{1}{c}{$34\pm2$} \\
  Untargeted & \multicolumn{1}{c}{$30^{+6}_{-5}$} & \multicolumn{1}{c}{$34^{+7}_{-6}$} & \multicolumn{1}{c}{$25^{+8}_{-7}$} & \multicolumn{1}{c}{$30\pm4$} \\
  \hline
  \end{tabular}
  \parbox{\hsize}{\emph{Notes:} The uncertainties on $f_{\rm HV}$ correspond to binomial confidence intervals
                  computed following the approach of \citet{2011PASA...28..128C}.}
  \end{minipage}
\end{table}

It is important to note that $f_{\rm HV}$,
defined as the fraction of SNe~Ia with $V_{\rm Si} \geq 12000$ km~s$^{-1}$, remains statistically consistent
across the different morphological classes of the host galaxies in our sample (Table~\ref{tabSNhostmorphfHV}).
We further examine the distribution of $f_{\rm HV}$ values within different morphological bins by separating
SNe~Ia discovered in targeted and untargeted surveys.
In both cases, the HV fractions remain consistent, within uncertainties, across a broad range of host morphologies,
from E--S0 systems to early- and late-type spirals.
This conclusion is further supported by Fisher's exact and Barnard's tests,
which reveal no statistically significant differences in the distribution of HV and NV events among the various
host galaxy morphological bins ($P>0.1$ in all cases).

The absence of a significant variation in $f_{\rm HV}$ with morphology provides no evidence that the relative occurrence of the HV and NV subgroups is strongly governed by the global host galaxy properties traced by morphology. In particular, the lack of a significantly enhanced HV fraction in early-type, actively star-forming galaxies does not support a simple scenario in which HV SNe~Ia originate predominantly from younger and/or more metal-rich progenitor populations. However, morphology is not a direct measurement of stellar age, metallicity, or star-formation activity, and these results do not exclude differences in the local environments of the progenitor systems.

\subsection{\mbox{Si\,{\sc ii}} velocity versus galactocentric distance}
\label{RESults2}

\begin{figure}
\centering
    \includegraphics[width=0.995\hsize]{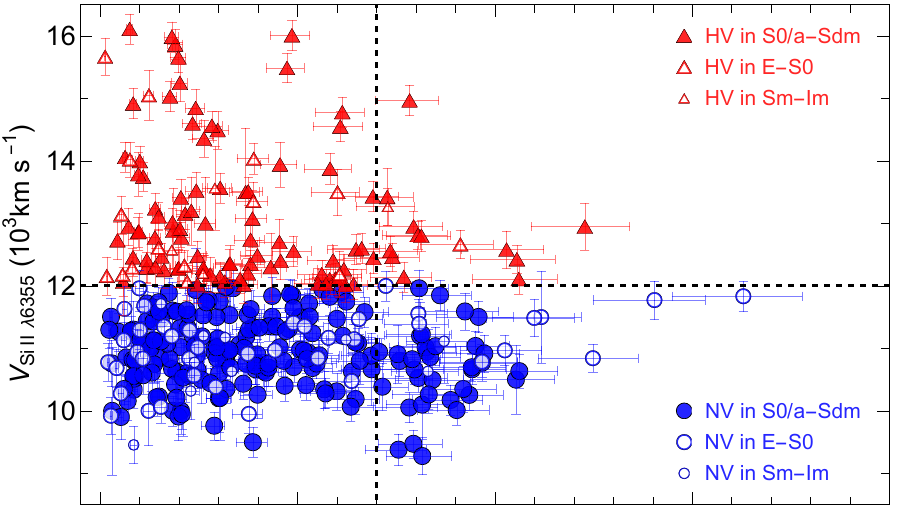}\\
    \vspace{-0.04cm}
    \hspace*{-0.15mm}\includegraphics[width=1\hsize]{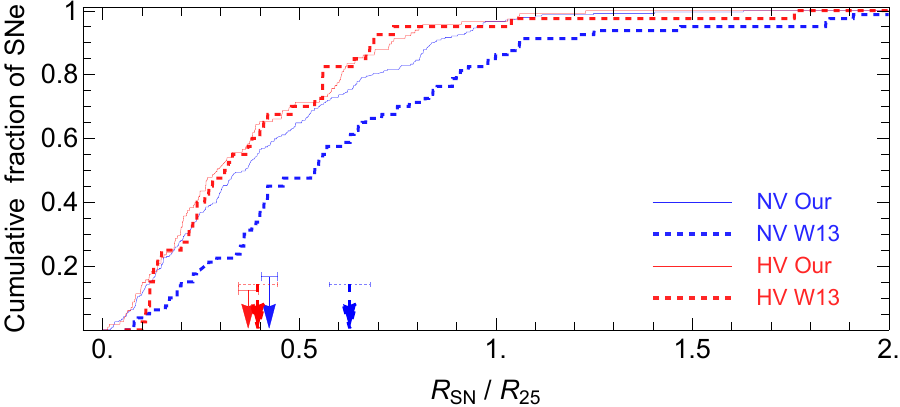}
    \caption{\emph{Upper panel:} distribution of \mbox{Si\,{\sc ii}} velocities as a function of
             the normalised galactocentric distance, $R_{\rm SN}/R_{25}$, for HV (red) and NV (blue) SNe~Ia,
             adopting the inclination correction scheme of \citetalias{2013Sci...340..170W} (see text for details).
             SNe~Ia in S0/a--Sdm and E--S0 hosts are indicated by filled and open symbols,
             respectively, while those in Sm--Im galaxies are shown by smaller open symbols.
             The horizontal dashed line indicates the division between HV and NV events.
             The vertical dashed line indicates $R_{\rm SN}/R_{25}=0.7$, the boundary adopted by \citetalias{2013Sci...340..170W}
             to highlight the central concentration of HV events.
             \emph{Bottom panel:} cumulative $R_{\rm SN}/R_{25}$ distributions.
             Dashed red and blue curves represent the HV and NV subsamples of \citetalias{2013Sci...340..170W},
             respectively, while the solid curves correspond to our data.
             Arrows indicate the mean $R_{\rm SN}/R_{25}$ of each subgroup,
             with error bars denoting the standard errors of the means.}
    \label{VSiRSNR25}
\end{figure}
\begin{table*}
  \centering
  \begin{minipage}{175mm}
  \caption{Comparison of the radial distributions between the different subgroups of SNe~Ia.}
  \tabcolsep 2.85pt
  \label{RSNR25}
    \begin{tabular}{ccccccccccrrrr}
    \hline
  \multicolumn{1}{c}{Morph} & \multicolumn{4}{c}{------------------ Subsample~1 ------------------} & \multicolumn{1}{c}{vs} & \multicolumn{4}{c}{------------------ Subsample~2 ------------------} & \multicolumn{1}{c}{$P_{\rm KS}^{\rm MC}$} & \multicolumn{1}{c}{$P_{\rm AD}^{\rm MC}$} & \multicolumn{1}{c}{$\mathrm{Med}(P_{\rm KS})$} & \multicolumn{1}{c}{$\mathrm{Med}(P_{\rm AD})$}\\
  & \multicolumn{1}{c}{Un/Targeted} & \multicolumn{1}{c}{SN} & \multicolumn{1}{c}{$N_{\rm SN}$} & \multicolumn{1}{c}{$\langle R_{\rm SN}/R_{25} \rangle$} && \multicolumn{1}{c}{Un/Targeted} & \multicolumn{1}{c}{SN} & \multicolumn{1}{c}{$N_{\rm SN}$} & \multicolumn{1}{c}{$\langle R_{\rm SN}/R_{25} \rangle$} && && \\
  \hline
     All & All & NV (our) & 239 & 0.42$\pm$0.02 & vs & All & HV (our) & 115 & 0.37$\pm$0.02 & 0.343 & 0.181 & $0.281^{+0.127}_{-0.110}$ & $0.193^{+0.101}_{-0.071}$\\
     All & All & All (our) & 88 & 0.45$\pm$0.03 & vs & All & All (\citetalias{2013Sci...340..170W}) & 88 & 0.56$\pm$0.04 & 0.260 & 0.145 & $0.293^{+0.095}_{-0.020}$ & $0.143^{+0.022}_{-0.020}$\\
     All & All & NV (\citetalias{2013Sci...340..170W}) & 74 & 0.61$\pm$0.05 & vs & All & HV (\citetalias{2013Sci...340..170W}) & 38 & 0.41$\pm$0.05 & \textbf{0.013} & \textbf{0.003} & $\textbf{0.018}^{+\textbf{0.011}}_{-\textbf{0.007}}$ & $\textbf{0.004}^{+\textbf{0.003}}_{-\textbf{0.001}}$\\
     All & Targeted & All (our) & 211 & 0.44$\pm$0.02 & vs & Untargeted & All (our) & 143 & 0.35$\pm$0.03 & $<$\textbf{0.001} & $<$\textbf{0.001} & $<$\textbf{0.001} & $<$\textbf{0.001}\\
     All & Untargeted & NV (our) & 100 & 0.36$\pm$0.03 & vs & Untargeted & HV (our) & 43 & 0.35$\pm$0.05 & 0.495 & 0.714 & $0.562^{+0.212}_{-0.217}$ & $0.691^{+0.118}_{-0.133}$\\
     All & Targeted & NV (our) & 139 & 0.47$\pm$0.02 & vs & Targeted & HV (our) & 72 & 0.38$\pm$0.03 & 0.138 & \textbf{0.039} & $0.235^{+0.103}_{-0.077}$ & $0.062^{+0.045}_{-0.025}$\\
     \\
     Elliptical & All & NV (our) & 32 & 0.43$\pm$0.08 & vs & All & HV (our) & 12 & 0.27$\pm$0.08 & 0.603 & 0.291 & $0.672^{+0.251}_{-0.203}$ & $0.342^{+0.139}_{-0.141}$\\
     Elliptical & Targeted & All (our) & 19 & 0.56$\pm$0.09 & vs & Untargeted & All (our) & 25 & 0.25$\pm$0.07 & \textbf{0.001} & \textbf{0.003} & $\textbf{0.003}^{+\textbf{0.002}}_{-\textbf{0.001}}$ & $\textbf{0.002}^{+\textbf{0.002}}_{-\textbf{0.001}}$\\
     \\
     S0/a--Sdm$^*$ & All & NV (our) & 151 & 0.43$\pm$0.02 & vs & All & HV (our) & 77 & 0.42$\pm$0.03 & 0.840 & 0.828 & $0.834^{+0.102}_{-0.197}$ & $0.797^{+0.090}_{-0.122}$\\
     S0/a--Sdm$^*$ & Targeted & All (our) & 149 & 0.44$\pm$0.02 & vs & Untargeted & All (our) & 79 & 0.40$\pm$0.03 & 0.108 & \textbf{0.049} & $0.127^{+0.028}_{-0.026}$ & $\textbf{0.046}^{+\textbf{0.006}}_{-\textbf{0.006}}$\\
     S0/a--Sdm$^*$ & Untargeted & NV (our) & 57 & 0.37$\pm$0.04 & vs & Untargeted & HV (our) & 22 & 0.49$\pm$0.08 & 0.264 & 0.165 & $0.421^{+0.125}_{-0.128}$ & $0.166^{+0.039}_{-0.029}$\\
     S0/a--Sdm$^*$ & Targeted & NV (our) & 94 & 0.47$\pm$0.03 & vs & Targeted & HV (our) & 55 & 0.40$\pm$0.03 & 0.281 & 0.126 & $0.364^{+0.145}_{-0.111}$ & $0.139^{+0.079}_{-0.047}$\\
  \hline
  \end{tabular}
  \parbox{\hsize}{\emph{Notes:} $P_{\rm KS}^{\rm MC}$ and $P_{\rm AD}^{\rm MC}$ are bootstrap MC probabilities derived from
                  $10^5$ resamplings of the compared subsamples using the two-sample KS and AD tests.
                  The final two columns list the median $P$-values from $10^4$ uncertainty-perturbation
                  realisations, in which the individual \mbox{Si\,{\sc ii}} velocities and normalised galactocentric distances were perturbed
                  according to their uncertainties, with the 16th--84th percentile range.
                  Mean values of $R_{\rm SN}/R_{25}$ and their standard errors are listed for each subsample.
                  Statistically significant differences $(P\leq0.05)$ between the distributions are highlighted in bold.
                  S0/a--Sdm subsamples marked with an asterisk $(*)$ correspond to host galaxies with inclinations $i < 70^{\circ}$.}
  \end{minipage}
\end{table*}

In the upper panel of Fig.~\ref{VSiRSNR25}, we show the distribution of \mbox{Si\,{\sc ii}} velocities as a function of
the normalised galactocentric distance, $R_{\rm SN}/R_{25}$, for SNe~Ia in our sample.
The bottom panel shows the CDFs of $R_{\rm SN}/R_{25}$ for NV and HV SNe~Ia in both our sample and that of \citetalias{2013Sci...340..170W}.
Statistical comparisons between the $R_{\rm SN}/R_{25}$ distributions of the different subsamples,
based on two-sample KS and AD tests, are summarised in Table~\ref{RSNR25}.
To propagate measurement uncertainties, we performed $10^4$ MC realisations in which the individual
\mbox{Si\,{\sc ii}} velocities and normalised galactocentric distances were perturbed according to Gaussian distributions
defined by their uncertainties. The median KS and AD $P$-values, together with their 16th--84th percentile ranges,
are presented in the final two columns of Table~\ref{RSNR25}.
Only subsamples containing more than 10 objects are included in the statistical analysis to avoid unreliable results
caused by small-number statistics.

First, in Table~\ref{RSNR25} we compare the radial distributions of NV and HV SNe~Ia for the full sample of 354 events,
adopting an inclination correction scheme following that of \citetalias{2013Sci...340..170W}.
Our sample shows no statistically significant difference between the radial distributions of the two velocity subgroups.
Although HV SNe~Ia exhibit a slightly smaller mean galactocentric distance than NV events (see Table~\ref{RSNR25} and Fig.~\ref{VSiRSNR25}),
the KS and AD tests provide no statistically significant evidence for a difference between the two distributions.
Therefore, our data provide no statistically significant evidence that HV SNe~Ia are more centrally concentrated than NV events,
in contrast to the findings of \citetalias{2013Sci...340..170W}.
This conclusion is also evident from the corresponding CDFs shown in the bottom panel of Fig.~\ref{VSiRSNR25}.

One possible explanation for the discrepancy between our results and those of \citetalias{2013Sci...340..170W}
is the presence of systematic differences in the measurements of SN galactocentric distances.
To test this possibility, we directly compared our $R_{\rm SN}/R_{25}$ measurements with those reported by \citetalias{2013Sci...340..170W}
for the 88 SNe~Ia common to both samples.
The KS and AD tests provide no statistically significant evidence for a difference between the resulting radial distributions (see the second row of Table~\ref{RSNR25}).
Thus, this comparison provides no statistically significant evidence that differences in the determination of galactocentric distances account for the discrepant conclusions; however, it does not exclude a contribution from differences in data sources and measurement procedures.
For completeness, an object-by-object comparison of the normalised galactocentric distances for
the 88 SNe~Ia common to \citetalias{2013Sci...340..170W} is presented in Appendix~\ref{app:paired}.

We then applied our disturbance criteria to the \citetalias{2013Sci...340..170W} sample,
removing host galaxies exhibiting prominent tidal interactions, ongoing mergers, or well-defined post-merger features.
After these exclusions, the refined \citetalias{2013Sci...340..170W} sample comprised 74 NV and 38 HV SNe~Ia.
A comparison of their radial distributions (third row of Table~\ref{RSNR25}) still reveals a statistically significant difference
between the two velocity subgroups, thereby confirming the robustness of the original result reported by \citetalias{2013Sci...340..170W}.
Consequently, the discrepancy between our findings and those of \citetalias{2013Sci...340..170W} is not readily explained solely by differences in the determination of $R_{\rm SN}/R_{25}$ or by the inclusion of morphologically disturbed host galaxies.
Other factors, including survey-selection effects, are investigated below.

To further study the origin of the discrepancies, we examined the discovery channels of the SNe in our sample,
distinguishing between galaxy-targeted and untargeted surveys.
The \citetalias{2013Sci...340..170W} sample is composed entirely of SNe~Ia discovered by targeted surveys,
with no contribution from untargeted searches.
By contrast, our sample, although still dominated by 211 events from targeted surveys ($\sim60$ per cent),
also includes 143 SNe~Ia discovered by untargeted surveys ($\sim40$ per cent).

In the fourth row of Table~\ref{RSNR25},
we show that the radial distributions of SNe~Ia differ significantly between
events discovered by targeted $(\langle R_{\rm SN}/R_{25} \rangle = 0.44\pm0.02)$
and untargeted $(\langle R_{\rm SN}/R_{25} \rangle = 0.35\pm0.03)$ surveys.
We therefore subdivide both samples into NV and HV subgroups (see next rows in Table~\ref{RSNR25}).
For the untargeted sample, the KS and AD tests provide no statistically significant evidence for a difference between the radial distributions of NV and HV SNe~Ia.
By contrast, within the targeted sample, the direct AD test yields a marginally significant difference between the radial distributions of NV and HV SNe~Ia
($P_{\rm AD}^{\rm MC}=0.039$), in qualitative agreement with the results of \citetalias{2013Sci...340..170W}. However, after propagating the measurement uncertainties,
the median AD probability becomes $0.062^{+0.045}_{-0.025}$, just above the adopted significance threshold.
The evidence for this difference should therefore be regarded as marginal.

These findings strongly suggest that the apparent central concentration of HV SNe~Ia reported by \citetalias{2013Sci...340..170W} is not a universal property of the HV population, but instead depends sensitively on the manner in which the SN sample is assembled.
This interpretation is consistent with recent analyses based on untargeted samples,
which generally find weaker environmental distinctions between NV and HV SNe~Ia
\citep[e.g.][]{2025A&A...698A.305M,2025A&A...694A..13B,2026A&A...705A..76B}.

Targeted surveys are inherently biased toward luminous and morphologically prominent galaxies,
which dominate preselected monitoring programmes \citep[e.g.][]{1999A&A...351..459C,2011MNRAS.412.1441L}.
Consequently, they preferentially sample massive, metal-rich host galaxies
\citep[e.g.][]{2016A&A...591A..48G}, together with their characteristic surface-brightness
distributions and radial stellar-population gradients, rather than the broader range of
environments represented in untargeted wide-field surveys such as ASAS-SN and ZTF
\citep[e.g.][]{2017PASP..129j4502K,2025A&A...694A...1R}.
Such biases can in principle alter both the Gaussian-model weights of the NV and HV populations in the \mbox{Si\,{\sc ii}} velocity distribution (see Sect.~\ref{RESults1})
and their inferred spatial distributions within host galaxies (Table~\ref{RSNR25}).
In contrast, untargeted surveys provide a more complete and less environmentally biased sampling of the local SN~Ia population.
The disappearance of the NV--HV radial distribution differences in the untargeted sample therefore argues against
a strong intrinsic environmental segregation between the two velocity subgroups and suggests that any genuine physical differences are
likely weaker than originally inferred from targeted samples alone.

Moreover, motivated by our previous studies of SN radial distributions in nearby galaxies
\citep[e.g.][]{2016MNRAS.456.2848H,2019MNRAS.490..718B,2021MNRAS.505L..52H},
we further investigate the spatial distributions of SNe~Ia separately for different host galaxy morphologies.
As mentioned above, for events occurring in elliptical galaxies, we use projected galactocentric distances,
as no physically well-motivated deprojection can be applied to these spheroid-dominated systems \citep[e.g.][]{2019MNRAS.490..718B}.
By contrast, for spiral galaxies, which comprise both bulge and disc stellar components, we apply inclination corrections
under the assumption that the majority of SNe~Ia originate in the galactic discs rather than in the bulges \citep[e.g.][]{2016MNRAS.456.2848H}.
To minimise the effects of projection uncertainties and internal extinction, the analysis of spiral hosts is restricted to systems
with $i < 70^{\circ}$.
Lenticular (S0) galaxies are not considered separately because the available number of SNe~Ia is insufficient for a statistically meaningful
comparison (Table~\ref{tabSNhostmorph}).
In addition, the mixed bulge--disc nature of these systems complicates the interpretation of deprojected galactocentric distances
\citep[e.g.][]{2016MNRAS.456.2848H}.

In elliptical hosts, we again find no statistically significant difference between the radial distributions of
NV and HV SNe~Ia (Table~\ref{RSNR25}).
Similarly, for the S0/a--Sdm subsample, the KS and AD tests provide no statistically significant evidence for a difference between the radial distributions of NV and HV SNe~Ia,
and the mean galactocentric distances of the two velocity subgroups are nearly identical.
This result is particularly important because the majority of SNe~Ia in our sample occur in
these host galaxies (see Table~\ref{tabSNhostmorph}) and because the original claim of a stronger central concentration of
HV events was primarily associated with disc-dominated systems.
Such systems are known to exhibit negative radial metallicity gradients,
implying systematically higher metallicities toward their central regions \citep[e.g.][]{2015A&A...581A.103G}.
Therefore, if HV SNe~Ia were preferentially associated with more metal-rich younger progenitor environments (\citetalias{2013Sci...340..170W}),
a stronger central concentration would naturally be expected.
However, our results do not support such a trend.

We further compare the radial distributions of SNe~Ia discovered by targeted and untargeted surveys within each morphological class.
For elliptical hosts, the difference between the targeted and untargeted samples remains highly significant,
while for S0/a--Sdm galaxies it is present only at a marginal significance level (Table~\ref{RSNR25}).
A plausible explanation is that untargeted searches detect many SNe~Ia in small elliptical galaxies (Tables~\ref{GLsizes} and \ref{Masses}), whose stellar-population radial distributions may differ from those of larger ellipticals. Such small host galaxies are generally not included, or are strongly under-represented, in the galaxy catalogues monitored by targeted SN surveys \citep[e.g.][]{2011MNRAS.412.1441L,2011MNRAS.412.1473L}.
We then subdivide the targeted and untargeted samples into NV and HV events for S0/a--Sdm hosts.
In all cases, the radial distributions of the two velocity subgroups remain statistically consistent.
In particular, even within the targeted S0/a--Sdm subsample,
we find no significant difference between the NV and HV distributions.
This likely reflects both the reduced statistical power after subdivision into smaller samples and
the intrinsically weak dependence of SN~Ia ejecta velocity on galactocentric position once host-galaxy morphology is controlled.

To assess the robustness of our results against the adopted velocity threshold separating NV and HV SNe~Ia,
we repeat the analysis presented in this section using threshold values between 11500 and 12500 km~s$^{-1}$ in steps of 100 km~s$^{-1}$.
The resulting radial distributions and associated statistical tests remain essentially unchanged across
the entire range of thresholds considered.
We therefore conclude that the results reported in Table~\ref{RSNR25} are not driven by
the specific choice of the 12000 km~s$^{-1}$ dividing velocity.

Another potential source of systematic uncertainty is that our velocity compilation combines measurements from several literature sources
(see Sect.~\ref{Iadata}) employing slightly different procedures to determine the \mbox{Si\,{\sc ii}}~$\lambda$6355 absorption velocity.
However, a recent comparison of the most commonly adopted methodologies, including Gaussian-profile fitting and measurements based on smoothed spectra,
showed that they yield highly consistent velocity estimates with no significant systematic differences \citep{2021ApJ...923..267D}.

As a robustness check, we repeat the galactocentric distance analysis using only the homogeneous velocity measurements of \citet{2021ApJ...923..267D}.
Their sample contains 111 SNe~Ia in common with ours, of which 110 belong to our targeted subsample, with an HV fraction of $25^{+5}_{-4}$ per cent,
close to that of our targeted sample (Table~\ref{tabSNhostmorphfHV}).
Comparison of the 82 NV and 28 HV SNe~Ia yields KS and AD $P$-values below 0.04, reproducing the statistically significant central concentration of
HV events reported by \citetalias{2013Sci...340..170W} and recovered for our targeted subsample.

We perform the same analysis using the homogeneous velocity measurements of \citet{2025A&A...694A...9B},
based entirely on untargeted SN discoveries. We identify 98 SNe~Ia in common with our sample, comprising 71 NV and 27 HV events,
with an HV fraction of $25^{+5}_{-5}$ per cent.
In this case, the KS and AD tests reveal no statistically significant differences between the galactocentric distance distributions of the NV and HV events,
consistent with our untargeted subsample.
These two independent homogeneous datasets therefore reproduce the contrasting behaviours of our targeted and untargeted samples,
demonstrating that this dichotomy is robust against differences in \mbox{Si\,{\sc ii}} velocity measurement methodology and
is unlikely to arise from the heterogeneous compilation of literature velocities.

Thus, for our sample, comprising $\sim60$ per cent targeted and $\sim40$ per cent untargeted events,
no statistically significant NV and HV difference in the radial distributions is detected within any of the host galaxy morphological subsamples examined.
This lack of significant environmental differences weakens the interpretation that the observed \mbox{Si\,{\sc ii}} velocity
diversity primarily reflects distinct progenitor populations occupying systematically different galactic environments
(e.g. \citetalias{2013Sci...340..170W}; \citealt{2015MNRAS.446..354P,2023arXiv230410601N}).
Instead, the results favour a scenario in which a substantial fraction of the observed velocity diversity arises from intrinsic explosion asymmetries and viewing-angle effects, while not excluding a possible contribution from local progenitor conditions that are not directly measured in this work \citep[e.g.][]{2024MNRAS.531.1988L,2026arXiv260622173G}.
This interpretation is consistent with theoretical predictions \citep[e.g.][]{2007Sci...315..825M,2010Natur.466...82M,2013MNRAS.429.1156S}
and recent observational studies \citep[e.g.][]{2020MNRAS.499.5325Z,2025A&A...698A.305M,2025A&A...694A..13B}.

\begin{figure*}
\centering
    \includegraphics[width=0.5\hsize]{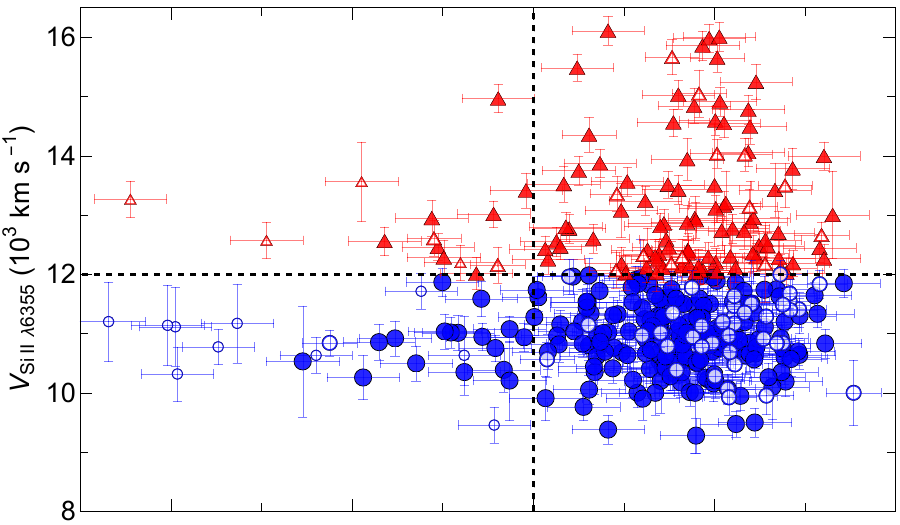}\\
    \vspace{-0.09cm}
    \hspace*{21.9mm}\raisebox{1.05mm}{\includegraphics[width=0.5\hsize]{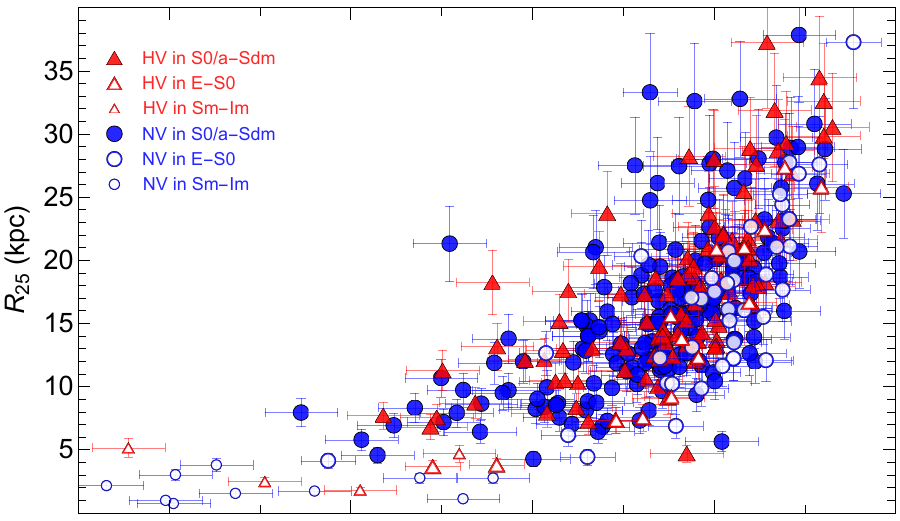}}\hspace*{0.6mm}\raisebox{-3.4mm}{\includegraphics[width=0.118\hsize]{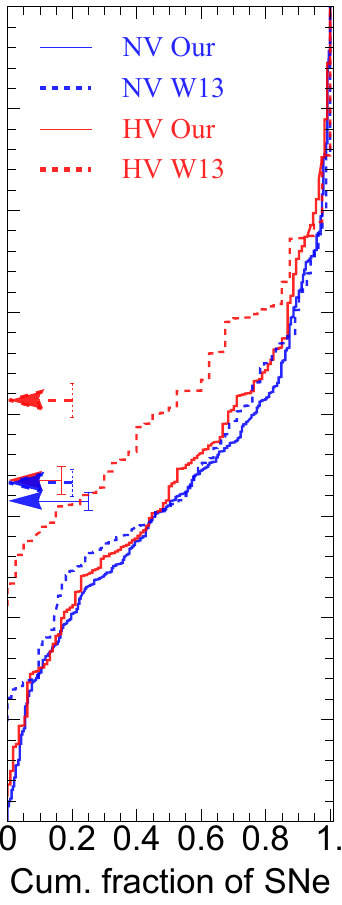}}\\
    \vspace{-0.48cm}
    \hspace*{1.3mm}\includegraphics[width=0.509\hsize]{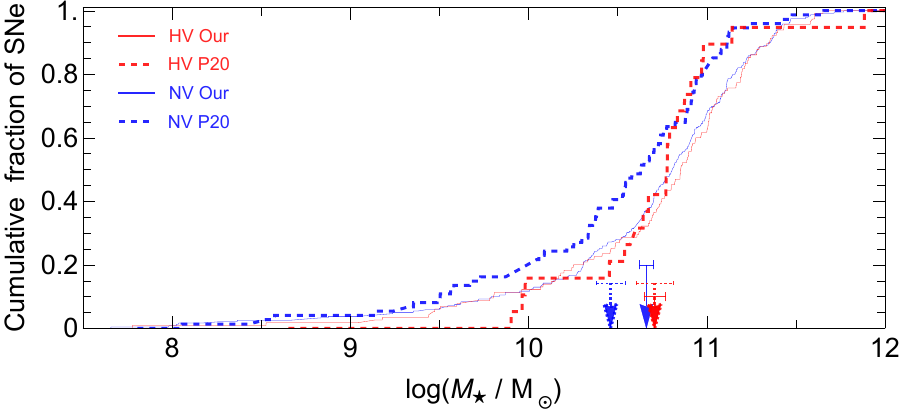}
    \caption{\emph{Upper panel:} distribution of \mbox{Si\,{\sc ii}} velocities as a function of host galaxy stellar mass,
             $\log(M_{\ast}/{\rm M_{\odot}})$, for SNe~Ia. The symbols, colours, and horizontal dashed line are the same as
             in Fig.~\ref{VSiRSNR25}. The vertical dashed line marks $\log(M_{\ast}/{\rm M_{\odot}})=10$,
             adopted by \citetalias{2020ApJ...895L...5P} to separate low- and high-mass host galaxies.
             \emph{Middle left panel:} host galaxy size, $R_{25}$, as a function of stellar mass for NV and HV SNe~Ia.
             \emph{Middle right panel:} cumulative distributions of the host galaxy sizes, $R_{25}$,
             in our sample (solid curves) and that of \citetalias{2013Sci...340..170W} (dashed curves).
             \emph{Bottom panel:} cumulative distributions of host galaxy stellar masses for SNe~Ia in
             our sample (solid curves) and the redshift-matched ($z\leq0.04$) subsample of
             \citetalias{2020ApJ...895L...5P} (dashed curves).
             Arrows indicate the mean values of $R_{25}$ (middle right panel) and $\log(M_{\ast}/{\rm M_{\odot}})$ (bottom panel),
             with error bars denoting the standard errors of the means.}
    \label{VSiMassPLcdf}
\end{figure*}

\subsection{\mbox{Si\,{\sc ii}} velocity versus host galaxy size and stellar mass}
\label{RESults3}

\begin{table*}
  \centering
  \begin{minipage}{170mm}
  \caption{Comparison of the host galaxy sizes ($R_{25}$ in kpc) between the different subgroups of SNe~Ia.}
  \tabcolsep 2.5pt
  \label{GLsizes}
    \begin{tabular}{ccccccccccrrrr}
    \hline
  \multicolumn{1}{c}{Morph} & \multicolumn{4}{c}{------------------ Subsample~1 ------------------} & \multicolumn{1}{c}{vs} & \multicolumn{4}{c}{------------------ Subsample~2 ------------------} & \multicolumn{1}{c}{$P_{\rm KS}^{\rm MC}$} & \multicolumn{1}{c}{$P_{\rm AD}^{\rm MC}$} & \multicolumn{1}{c}{$\mathrm{Med}(P_{\rm KS})$} & \multicolumn{1}{c}{$\mathrm{Med}(P_{\rm AD})$}\\
  & \multicolumn{1}{c}{Un/Targeted} & \multicolumn{1}{c}{SN} & \multicolumn{1}{c}{$N_{\rm SN}$} & \multicolumn{1}{c}{$\langle R_{25} \rangle$} && \multicolumn{1}{c}{Un/Targeted} & \multicolumn{1}{c}{SN} & \multicolumn{1}{c}{$N_{\rm SN}$} & \multicolumn{1}{c}{$\langle R_{25} \rangle$} && && \\
  \hline
     All & All & NV (our) & 239 & 15.7$\pm$0.4 & vs & All & HV (our) & 115 & 16.7$\pm$0.7 & 0.308 & 0.349 & $0.356^{+0.181}_{-0.141}$ & $0.365^{+0.179}_{-0.140}$\\
     All$^\dag$ & All & All (our) & 88 & 19.2$\pm$0.7 & vs & All & All (\citetalias{2013Sci...340..170W}) & 88 & 17.6$\pm$0.7 & 0.206 & 0.179 & $0.216^{+0.077}_{-0.061}$ & $0.170^{+0.039}_{-0.034}$\\
     All & Targeted & All (our) & 211 & 18.0$\pm$0.5 & vs & Untargeted & All (our) & 143 & 13.1$\pm$0.5 & $<$\textbf{0.001} & $<$\textbf{0.001} & $<$\textbf{0.001} & $<$\textbf{0.001}\\
     All & Untargeted & NV (our) & 100 & 13.2$\pm$0.6 & vs & Untargeted & HV (our) & 43 & 12.8$\pm$1.0 & 0.588 & 0.820 & $0.592^{+0.176}_{-0.191}$ & $0.717^{+0.138}_{-0.179}$\\
     All & Targeted & NV (our) & 139 & 17.5$\pm$0.6 & vs & Targeted & HV (our) & 72 & 19.1$\pm$0.8 & 0.077 & 0.087 & $0.103^{+0.078}_{-0.052}$ & $0.107^{+0.078}_{-0.046}$\\
     \\
     Elliptical & All & NV (our) & 32 & 17.1$\pm$1.3 & vs & All & HV (our) & 12 & 15.5$\pm$2.4 & 0.729 & 0.584 & $0.621^{+0.103}_{-0.331}$ & $0.385^{+0.114}_{-0.177}$\\
     Elliptical & Targeted & All (our) & 19 & 20.5$\pm$1.7 & vs & Untargeted & All (our) & 25 & 13.8$\pm$1.2 & \textbf{0.008} & \textbf{0.003} & $\textbf{0.010}^{+\textbf{0.002}}_{-\textbf{0.006}}$ & $\textbf{0.002}^{+\textbf{0.001}}_{-\textbf{0.001}}$\\
     \\
     S0/a--Sdm & All & NV (our) & 189 & 16.2$\pm$0.5 & vs & All & HV (our) & 96 & 17.7$\pm$0.7 & 0.201 & 0.113 & $0.269^{+0.144}_{-0.110}$ & $0.178^{+0.124}_{-0.077}$\\
     S0/a--Sdm & Targeted & All (our) & 184 & 18.0$\pm$0.5 & vs & Untargeted & All (our) & 101 & 14.3$\pm$0.6 & \textbf{0.001} & $<$\textbf{0.001} & $<$\textbf{0.001} & $<$\textbf{0.001}\\
     S0/a--Sdm & Untargeted & NV (our) & 70 & 14.2$\pm$0.7 & vs & Untargeted & HV (our) & 31 & 14.6$\pm$1.0 & 0.716 & 0.940 & $0.795^{+0.125}_{-0.212}$ & $0.917^{+0.053}_{-0.108}$\\
     S0/a--Sdm & Targeted & NV (our) & 119 & 17.4$\pm$0.6 & vs & Targeted & HV (our) & 65 & 19.2$\pm$0.8 & 0.100 & 0.095 & $0.177^{+0.110}_{-0.081}$ & $0.126^{+0.088}_{-0.052}$\\
  \hline
  \end{tabular}
  \parbox{\hsize}{\emph{Note:} The explanation of the $P$-values is the same as in Table~\ref{RSNR25}.
                  Subsamples marked with a dagger $(\dag)$ correspond to host galaxy sizes (in kpc) derived from
                  uncorrected $R_{25}$ measurements (in arcsec), i.e. without applying corrections for
                  Galactic foreground or internal host galaxy extinction (see Sect.~\ref{Hostdata}),
                  to ensure a direct comparison with the measurements of \citetalias{2013Sci...340..170W}.}
  \end{minipage}
\end{table*}
\begin{table*}
  \centering
  \begin{minipage}{175mm}
  \caption{Comparison of the host galaxy stellar masses ($\log(M_{\ast}/{\rm M_{\odot}})$ in dex) between the different subgroups of SNe~Ia.}
  \tabcolsep 2.3pt
  \label{Masses}
    \begin{tabular}{ccccccccccrrrr}
    \hline
  \multicolumn{1}{c}{Morph} & \multicolumn{4}{c}{------------------ Subsample~1 ------------------} & \multicolumn{1}{c}{vs} & \multicolumn{4}{c}{------------------ Subsample~2 ------------------} & \multicolumn{1}{c}{$P_{\rm KS}^{\rm MC}$} & \multicolumn{1}{c}{$P_{\rm AD}^{\rm MC}$} & \multicolumn{1}{c}{$\mathrm{Med}(P_{\rm KS})$} & \multicolumn{1}{c}{$\mathrm{Med}(P_{\rm AD})$}\\
  & \multicolumn{1}{c}{Un/Targeted} & \multicolumn{1}{c}{SN} & \multicolumn{1}{c}{$N_{\rm SN}$} & \multicolumn{1}{c}{$\langle \log(M_{\ast}/{\rm M_{\odot}}) \rangle$} && \multicolumn{1}{c}{Un/Targeted} & \multicolumn{1}{c}{SN} & \multicolumn{1}{c}{$N_{\rm SN}$} & \multicolumn{1}{c}{$\langle \log(M_{\ast}/{\rm M_{\odot}}) \rangle$} && && \\
  \hline
     All & All & NV (our) & 239 & 10.55$\pm$0.04 & vs & All & HV (our) & 115 & 10.71$\pm$0.06 & 0.647 & 0.866 & $0.730^{+0.200}_{-0.270}$ & $0.865^{+0.106}_{-0.213}$\\
     All & All & All (our) & 44 & 10.76$\pm$0.08 & vs & All & All (\citetalias{2020ApJ...895L...5P}) & 44 & 10.59$\pm$0.06 & 0.208 & 0.107 & $0.207^{+0.258}_{-0.079}$ & $0.123^{+0.084}_{-0.054}$\\
     All & Targeted & All (our) & 211 & 10.82$\pm$0.04 & vs & Untargeted & All (our) & 143 & 10.47$\pm$0.07 & $<$\textbf{0.001} & $<$\textbf{0.001} & $<$\textbf{0.001} & $<$\textbf{0.001}\\
     All & Untargeted & NV (our) & 100 & 10.47$\pm$0.09 & vs & Untargeted & HV (our) & 43 & 10.47$\pm$0.10 & 0.224 & 0.439 & $0.477^{+0.271}_{-0.230}$ & $0.552^{+0.159}_{-0.158}$\\
     All & Targeted & NV (our) & 139 & 10.80$\pm$0.04 & vs & Targeted & HV (our) & 72 & 10.85$\pm$0.07 & 0.089 & 0.262 & $0.169^{+0.163}_{-0.097}$ & $0.300^{+0.170}_{-0.122}$\\
     \\
     Elliptical & All & NV (our) & 32 & 11.06$\pm$0.09 & vs & All & HV (our) & 12 & 10.82$\pm$0.18 & 0.263 & 0.257 & $0.303^{+0.319}_{-0.173}$ & $0.275^{+0.299}_{-0.176}$\\
     Elliptical & Targeted & All (our) & 19 & 11.24$\pm$0.06 & vs & Untargeted & All (our) & 25 & 10.81$\pm$0.12 & \textbf{0.001} & \textbf{0.003} & $\textbf{0.008}^{+\textbf{0.013}}_{-\textbf{0.005}}$ & $\textbf{0.004}^{+\textbf{0.003}}_{-\textbf{0.002}}$\\
     \\
     S0/a--Sdm & All & NV (our) & 189 & 10.70$\pm$0.04 & vs & All & HV (our) & 96 & 10.78$\pm$0.05 & 0.072 & 0.168 & $0.181^{+0.174}_{-0.102}$ & $0.234^{+0.165}_{-0.106}$\\
     S0/a--Sdm & Targeted & All (our) & 184 & 10.80$\pm$0.03 & vs & Untargeted & All (our) & 101 & 10.58$\pm$0.06 & \textbf{0.043} & \textbf{0.004} & $\textbf{0.034}^{+\textbf{0.016}}_{-\textbf{0.013}}$ & $\textbf{0.005}^{+\textbf{0.002}}_{-\textbf{0.002}}$\\
     S0/a--Sdm & Untargeted & NV (our) & 70 & 10.58$\pm$0.07 & vs & Untargeted & HV (our) & 31 & 10.60$\pm$0.10 & 0.780 & 0.894 & $0.862^{+0.109}_{-0.231}$ & $0.894^{+0.069}_{-0.121}$\\
     S0/a--Sdm & Targeted & NV (our) & 119 & 10.70$\pm$0.04 & vs & Targeted & HV (our) & 65 & 10.86$\pm$0.06 & \textbf{0.023} & 0.097 & $0.077^{+0.098}_{-0.047}$ & $0.139^{+0.101}_{-0.064}$\\
  \hline
  \end{tabular}
  \parbox{\hsize}{\emph{Note:} The explanation of the $P$-values is the same as in Table~\ref{RSNR25}.}
  \end{minipage}
\end{table*}

To complement the analysis of SN~Ia galactocentric distances presented in Sect.~\ref{RESults2},
we now examine the relation between \mbox{Si\,{\sc ii}} velocity and global host galaxy properties.
While galactocentric distance provides an indirect positional diagnostic within the host galaxy, global properties such as galaxy size and stellar mass characterise the overall host population and are statistically related to its evolutionary and metal-enrichment history \citep[e.g.][]{2004ApJ...613..898T}.
We therefore investigate whether NV and HV SNe~Ia preferentially occur in galaxies of different physical sizes
($R_{25}$ in kpc) and stellar masses, $\log(M_{\ast}/{\rm M_{\odot}})$.
Such differences have been reported in several previous studies
(e.g. \citetalias{2013Sci...340..170W}; \citealt{2015MNRAS.446..354P}; \citetalias{2020ApJ...895L...5P}),
which found that HV SNe~Ia tend to occur in larger and more massive galaxies than their NV counterparts.
These findings were interpreted as evidence that the two velocity subgroups originate from progenitor populations
associated with systematically different host galaxy environments (metallicities).

Fig.~\ref{VSiMassPLcdf} shows the distributions of \mbox{Si\,{\sc ii}} velocities as a function of host galaxy stellar mass
(upper panel), host galaxy size versus stellar mass for NV and HV SNe~Ia (middle left panel), and the cumulative distributions
of $\log(M_{\ast}/{\rm M_{\odot}})$ for the two velocity subgroups (bottom panel),
comparing our sample with that of \citetalias{2020ApJ...895L...5P}.\footnote{It should be noted that
the \citetalias{2020ApJ...895L...5P} sample with available host stellar masses comprises 217 SNe~Ia
extending to $z<0.2$, substantially beyond the redshift limit of our sample ($z\leq0.04$).
To ensure a like-for-like comparison, we restrict the \citetalias{2020ApJ...895L...5P} sample to
the same redshift range, leaving 74 NV and 19 HV SNe~Ia with host stellar mass estimates.}
The middle right panel similarly compares the cumulative distributions of the host galaxy sizes, $R_{25}$ (kpc),
for our sample and the \citetalias{2013Sci...340..170W} sample.
The corresponding statistical comparisons are summarised in Tables~\ref{GLsizes} and \ref{Masses}.

First, we compare the host galaxy sizes and stellar masses of NV and HV SNe~Ia for the full sample.
As shown in Tables~\ref{GLsizes} and \ref{Masses}, we find no statistically significant differences between the two velocity subgroups.
The mean host galaxy sizes are $15.7\pm0.4$ and $16.7\pm0.7$ kpc for NV and HV SNe~Ia, respectively,
while the corresponding mean stellar masses are $\log(M_{\ast}/{\rm M_{\odot}})=10.55\pm0.04$ and $10.71\pm0.06$.
Although HV SNe~Ia tend to occur in slightly larger and more massive galaxies on average,
the KS and AD tests provide no statistically significant evidence for differences between the corresponding distributions.
Thus, in contrast to the results of \citetalias{2013Sci...340..170W} and \citetalias{2020ApJ...895L...5P}
\citep[see also][]{2015MNRAS.446..354P},
our full sample does not provide evidence that HV SNe~Ia preferentially reside in systematically larger or more massive host galaxies.
This is also evident from the corresponding CDFs shown in Fig.~\ref{VSiMassPLcdf}.
This principal conclusion concerning the absence of statistically significant NV/HV differences in host stellar masses remains unchanged when the comparison is restricted to 342 of the 354 SN~Ia host galaxies ($\sim97$ per cent of our sample) with REGALADE stellar mass measurements (Appendix~\ref{app:regalade-mass}).

As a consistency check analogous to that performed for the galactocentric distances in Sect.~\ref{RESults2},
we compare both our host galaxy size and stellar mass measurements with those reported by
\citetalias{2013Sci...340..170W} and \citetalias{2020ApJ...895L...5P}, respectively.
The comparison is based on the 88 SN~Ia hosts common to our sample and that of
\citetalias{2013Sci...340..170W} for the galaxy size analysis, and on the 44 host galaxies common to
our sample and that of \citetalias{2020ApJ...895L...5P} for the stellar mass analysis.
For both sets of common host galaxies, the KS and AD tests provide no statistically significant evidence of differences between the corresponding distributions
(second rows of Tables~\ref{GLsizes} and \ref{Masses}).
However, these results do not exclude systematic offsets arising from differences in the measurement procedures and calibrations.
The corresponding object-by-object offsets are quantified in Appendix~\ref{app:paired}.
The paired and distribution-level comparisons are complementary external checks and are not used as the basis for our physical conclusions.

An interesting picture emerges when the discovery strategy is considered.
The host galaxies of SNe~Ia discovered by targeted surveys are significantly larger and more massive
than those found by untargeted searches.
For the full sample, targeted hosts have a mean size of $18.0\pm0.5$ kpc and a mean stellar mass of $\log(M_{\ast}/{\rm M_{\odot}})=10.82\pm0.04$,
compared to $13.1\pm0.5$ kpc and $10.47\pm0.07$ for untargeted hosts.
These differences are highly significant according to both the KS and AD tests (Tables~\ref{GLsizes} and \ref{Masses}).
This result directly demonstrates that survey strategy introduces strong environmental selection effects,
preferentially sampling different host galaxy populations \citep[see, e.g.][]{2021ApJ...923..267D}.
Consequently, any comparison between NV and HV SNe~Ia that does not account for the targeted or untargeted nature
of the parent survey may be affected by systematic biases.

We next subdivide the targeted and untargeted samples into the NV and HV subgroups.
For the untargeted sample, neither the host galaxy sizes nor the stellar mass distributions
show statistically significant differences between NV and HV SNe~Ia
(Tables~\ref{GLsizes} and \ref{Masses}).
The mean host galaxy sizes of the two velocity subgroups are nearly identical
($13.2\pm0.6$ and $12.8\pm1.0$ kpc for NV and HV events, respectively),
while both populations have the same mean stellar mass,
$\log(M_{\ast}/{\rm M_{\odot}})=10.47$.
By contrast, within the targeted sample, HV SNe~Ia tend to occur in somewhat larger
and more massive host galaxies than NV events (see Tables~\ref{GLsizes} and \ref{Masses}).
Although the corresponding KS and AD tests for the phase-corrected velocities do not reach the adopted significance threshold,
the resulting $P$-values are systematically lower than those obtained for the untargeted sample, particularly for the host galaxy sizes.
Using the original uncorrected velocities, the targeted host-size comparison yields marginally significant median probabilities
($P_{\rm KS}=0.030$ and $P_{\rm AD}=0.040$; Appendix~\ref{app:uncorrected}), although their 16th--84th percentile ranges extend above the adopted threshold.
Thus, this tendency is sensitive to the velocity definition and should be interpreted cautiously.

When restricting the analysis to elliptical hosts, we again find no statistically significant differences
between the host galaxy sizes or stellar masses of NV and HV SNe~Ia.
Similarly, for the S0/a--Sdm subsample, which contains the majority of SNe~Ia in our sample (Table~\ref{tabSNhostmorph}),
no statistically significant differences are detected between the NV and HV distributions of either
$R_{25}$ or $\log(M_{\ast}/{\rm M_{\odot}})$, with only the KS test for the stellar mass comparison yielding
a marginal probability ($P_{\rm KS}=0.072$, see Table~\ref{Masses}).

We next compare the global properties of host galaxies for SNe~Ia discovered by targeted
and untargeted surveys within each morphological class.
For elliptical galaxies, the hosts of targeted discoveries are both significantly larger
and more massive than those found by untargeted surveys
($\log(M_{\ast}/{\rm M_{\odot}})=11.24\pm0.06$ versus $10.81\pm0.12$).
A similar difference is present for the full S0/a--Sdm sample. However, this particular stellar-mass comparison is sensitive to the inclusion of the 12 separately derived masses: the median KS probability changes from $0.034$ to $0.118$ and no longer reaches the adopted significance threshold when only REGALADE masses are used, although the corresponding median AD probability remains below this threshold ($0.026$; Appendix~\ref{app:regalade-mass}). This particular comparison should therefore be interpreted cautiously.

Finally, we subdivide the S0/a--Sdm sample according to the discovery strategy and compare
the host galaxy properties of the NV and HV subgroups.
For the untargeted sample, neither the host galaxy sizes nor the stellar masses differ
significantly between NV and HV SNe~Ia, with nearly identical mean values for both quantities.
Within the targeted sample, HV SNe~Ia again tend to occur in somewhat larger
and more massive host galaxies than NV events
($\log(M_{\ast}/{\rm M_{\odot}})=10.86\pm0.06$ versus $10.70\pm0.04$).
The direct statistical tests yield the lowest $P$-values of this section, particularly for the stellar mass comparison
($P_{\rm KS}^{\rm MC}=0.023$). However, after propagating the measurement uncertainties, the median KS probability is
$0.077^{+0.098}_{-0.047}$, above the adopted significance threshold. The AD test and the host-size comparison also remain above this threshold.
These results indicate that the apparent preference of HV SNe~Ia
for larger and more massive host galaxies is driven primarily by the targeted component
of the sample and becomes indistinguishable in the untargeted S0/a--Sdm sample.

As an additional robustness check, analogous to that performed in Sect.~\ref{RESults2},
we repeat the analysis using a range of velocity thresholds between 11500 and 12500 km~s$^{-1}$,
and the independent homogeneous \mbox{Si\,{\sc ii}} velocity measurements of
\citet[][for targeted discoveries]{2021ApJ...923..267D} and
\citet[][for untargeted discoveries]{2025A&A...694A...9B}.
The host galaxy size and stellar mass distributions, together with the corresponding statistical results,
remain qualitatively unchanged, demonstrating that the conclusions presented in this section
are robust against both the adopted NV/HV dividing velocity and the heterogeneous compilation
of literature velocity measurements.

Taken together with the analyses presented in Sects.~\ref{RESults1} and \ref{RESults2},
these findings weaken the interpretation that the observed \mbox{Si\,{\sc ii}} velocity diversity
is primarily driven by systematic differences in the global environments of SN~Ia progenitors
(e.g. \citetalias{2013Sci...340..170W}; \citealt{2015MNRAS.446..354P}; \citetalias{2020ApJ...895L...5P}).
Instead, our analysis shows that the previously reported associations between \mbox{Si\,{\sc ii}} velocity
and host galaxy size or stellar mass (metallicity) become substantially weaker once survey-selection effects
and host morphology are taken into account
\citep[see also][]{2021ApJ...923..267D,2025A&A...694A..10D,2025A&A...694A..13B}.
The absence of significant global environmental differences between the two velocity subgroups in
our sample favours a scenario in which a substantial fraction of the observed \mbox{Si\,{\sc ii}} velocity diversity
arises from intrinsic explosion properties, including ejecta asymmetries and viewing-angle effects,
while any environmental dependence is more likely linked to local progenitor conditions
than to the global characteristics of the host galaxy
\citep[e.g.][]{2007Sci...315..825M,2010Natur.466...82M,2013MNRAS.429.1156S,2018MNRAS.477.3567M,2020MNRAS.499.5325Z}.

\section{Conclusions}
\label{Concl}

We have carried out a comprehensive investigation of the environmental properties of 354 nearby ($z\leq0.04$)
spectroscopically normal SNe~Ia with published near-maximum-light \mbox{Si\,{\sc ii}}~$\lambda$6355 velocity measurements.
Combining our homogeneous determination of host morphology, SN galactocentric distance, host physical size,
and inferred stellar mass with a careful assessment of survey-selection effects,
we examined whether the NV and HV subgroups of nearby SN~Ia occupy systematically different galactic environments.

Our principal results are summarised as follows:

\begin{enumerate}[label=(\arabic*)]

\item
We confirm that the near-maximum-light \mbox{Si\,{\sc ii}}~$\lambda$6355 velocity distribution of spectroscopically normal SNe~Ia
is best described by two Gaussian components corresponding to the NV and HV populations (Fig.~\ref{Fits}).
For our full sample, the Gaussian-model weights are $73^{+2}_{-3}$ and $27^{+3}_{-2}$ per cent for the NV and HV components, respectively.
However, we demonstrate that these fractions depend on the discovery strategy.
The targeted subsample reproduces the relative proportions reported in previous targeted-survey studies,
whereas the untargeted subsample is dominated by the NV component and exhibits a substantially lower HV fraction.
This result indicates that the inferred relative occurrence of the two velocity populations is sensitive to
survey-selection effects \citep[e.g.][]{2024MNRAS.532.1887P} and suggests that part of the diversity reported in previous investigations
arises from observational biases rather than intrinsic differences within the SN~Ia population.

\item
We find no statistically significant dependence of the HV fraction, $f_{\rm HV}$,
defined as the fraction of SNe~Ia with $V_{\rm Si} \geq 12000$ km~s$^{-1}$, on host galaxy morphology (Table~\ref{tabSNhostmorphfHV}).
Although the mean \mbox{Si\,{\sc ii}} velocity exhibits a weak tendency to increase from late- to early-type spiral galaxies,
no statistically significant variation of $f_{\rm HV}$ is detected across the Hubble sequence.
In particular, the absence of a significant enhancement of the HV fraction in early-type, massive galaxies or
of a corresponding excess of NV events in late-type, actively star-forming systems, does not support
a simple scenario in which the observed ejecta velocity diversity is governed primarily by systematic differences in the global host galaxy properties associated with progenitor age and/or metallicity.

\item
We find no statistically significant differences between the galactocentric distance distributions of NV and HV SNe~Ia
(Fig.~\ref{VSiRSNR25} and Table~\ref{RSNR25}).
This result remains unchanged after adopting a methodology matching that of \citetalias{2013Sci...340..170W},
verifying the consistency of the galactocentric distances for the SNe~Ia common to our sample and that of \citetalias{2013Sci...340..170W},
separating host galaxies by morphology, and varying the velocity threshold defining the two subgroups.
The strongest environmental differences are instead found between SNe~Ia discovered by targeted and untargeted surveys,
consistent with survey-selection effects playing a major role in the previously reported central concentration of HV events.

\item
We find no statistically significant evidence that HV SNe~Ia occur preferentially in larger or more massive host galaxies than NV events
(Fig.~\ref{VSiMassPLcdf} and Tables~\ref{GLsizes}--\ref{Masses}).
Direct comparisons with the measurements of \citetalias{2013Sci...340..170W} and \citetalias{2020ApJ...895L...5P} for the SNe~Ia common to the samples provide external checks on the host galaxy size and stellar-mass measurements; however, they do not exclude the possibility that differences in data sources, measurement procedures, and calibrations contribute to the discrepant conclusions.
After accounting for survey-selection effects and host galaxy morphology,
the previously reported environmental trends become substantially weaker,
suggesting that global host galaxy properties alone are insufficient to explain
the observed \mbox{Si\,{\sc ii}} velocity diversity.

\item
Taken together, our analyses consistently show that the NV and HV subgroups do not occupy
systematically different global galactic environments.
The consistency of this result across multiple independent environmental diagnostics,
together with its robustness against different analysis methodologies and the adopted
NV/HV velocity threshold,
suggests that the environmental distinctions reported in earlier studies are not a general property of
the nearby SN~Ia population but are likely to be strongly influenced by the characteristics of the parent survey
\citep[e.g.][]{2021ApJ...923..267D,2025A&A...694A..13B}.

\item
These findings therefore challenge the interpretation that HV SNe~Ia arise predominantly from progenitor populations distinguished by the global host galaxy properties examined here.
Our results favour a scenario in which intrinsic explosion physics, including ejecta asymmetries and viewing-angle effects, produces a substantial fraction of the observed velocity diversity
\citep[e.g.][]{2007Sci...315..825M,2010Natur.466...82M,2013MNRAS.429.1156S,2018MNRAS.477.3567M,2020MNRAS.499.5325Z}. At the same time, the absence of significant differences in the global properties examined here does not exclude differences in local environments or progenitor properties \citep[e.g.][]{2024MNRAS.531.1988L,2026arXiv260622173G}.

\end{enumerate}

The main results presented in Sect.~\ref{RESults}, including the bimodal \mbox{Si\,{\sc ii}} velocity distribution and the environmental comparisons of NV and HV SNe~Ia, remain qualitatively unchanged when the original uncorrected velocities are used instead of those corrected following \citet{2011ApJ...742...89F} (see Appendix~\ref{app:uncorrected}).

Overall, our results indicate that much of the previously inferred environmental distinction
between NV and HV SNe~Ia originates from observational selection rather than intrinsic differences between the two populations.
Future progress will require large, homogeneous, predominantly untargeted SN~Ia samples,
such as those from ASAS-SN and ZTF \citep[e.g.][]{2017PASP..129j4502K,2025A&A...694A...1R},
combined with spatially resolved spectroscopy of explosion sites \citep[e.g.][]{2014A&A...572A..38G,2018ApJ...855..107G}
to disentangle the relative roles of local progenitor environments
and intrinsic explosion physics in shaping the observed ejecta velocity distribution.

\section*{Acknowledgements}

We thank the anonymous referee for their thoughtful comments
and constructive suggestions, which have helped us improve the paper.
The research was supported by the Higher Education and Science Committee of MESCS
RA (Research project \textnumero~24LCG--1C021).


\section*{Data Availability}

The data underlying this study are available as supplementary material accompanying the online version of the article.
The dataset contains, for each SN~Ia, the SN name, \mbox{Si\,{\sc ii}} velocity at maximum light, velocity measurement source,
survey type (targeted or untargeted), redshift, redshift source, normalised galactocentric distance,
host galaxy morphological $t$-type, host inclination, physical size (in kpc), host galaxy stellar mass (in dex), and, where available, SN~Ia $\Delta m_{15}(B)$ values together with their data sources.
The original \mbox{Si\,{\sc ii}} velocities with measurement phases are available in the references listed in the supplementary material.


\bibliographystyle{mnras}
\bibliography{Vreferences}


\section*{Supporting information}

Supplementary data are available at \emph{MNRAS} online.\\
\\
\textbf{supplementary.csv}\\
\\
Please note: Oxford University Press is not responsible for the
content or functionality of any supporting materials supplied by
the authors. Any queries (other than missing material) should be
directed to the corresponding author for the article.


\appendix

\section{Angular host galaxy radii as a function of redshift}
\label{app:angular-size}

\begin{figure}
\centering
\includegraphics[width=\columnwidth]{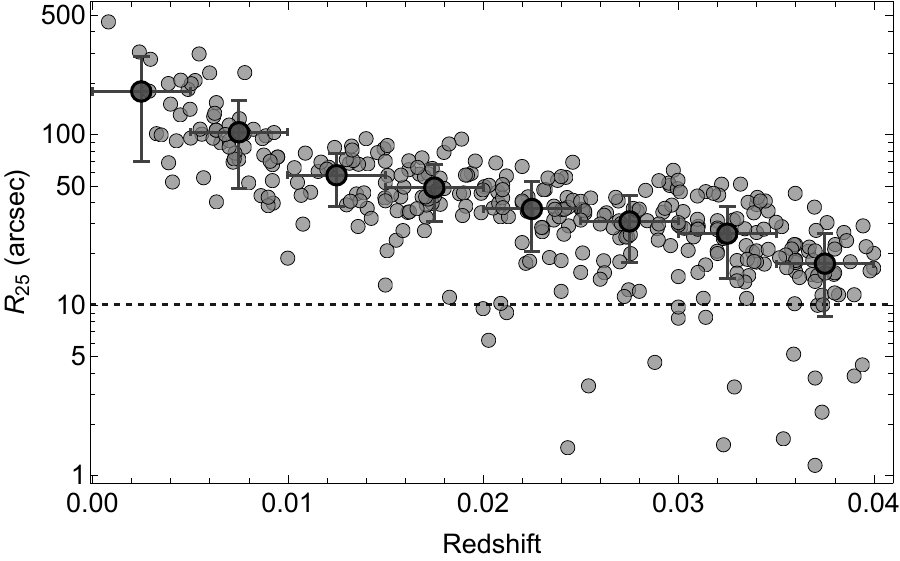}
\caption{Angular host galaxy radii, $R_{25}$, as a function of redshift for the 354 SNe~Ia in our sample.
The dashed line at $R_{25}=10$ arcsec is shown only as a visual reference.
The larger filled circles show the mean angular radii in eight redshift bins of width $\Delta z=0.005$; the vertical error bars are the corresponding standard deviations, and the horizontal error bars indicate the bin width.}
\label{fig:R25-redshift}
\end{figure}

To quantify the practical motivation for the adopted redshift limit, Fig.~\ref{fig:R25-redshift} shows the angular host galaxy radii, $R_{25}$, as a function of redshift for the 354 SNe~Ia in our sample.
The sample is divided into eight redshift bins of width $\Delta z=0.005$.
The mean angular radius decreases systematically from $\sim180$ arcsec in the $0\leq z<0.005$ bin to $\sim17$ arcsec in the highest-redshift bin, $0.035<z\leq0.04$.
The latter bin still contains 50 SNe~Ia. By comparison, adopting $z\leq0.035$ would reduce the sample from 354 to 304 events, while $z\leq0.02$ would retain only 144 events.

The adopted $z\leq0.04$ limit is a practical compromise. At the highest-redshift bin, the mean angular $R_{25}$ of $\sim17$ arcsec is sampled by approximately 43, 66, and 34 pixels in SDSS, Pan-STARRS, and SkyMapper, respectively. Thus, even near the adopted upper limit, the host galaxies remain sufficiently resolved for the geometrical measurements and morphological classification.
A stricter cut would increase the angular sizes of the hosts, but, as quantified above, would substantially reduce the available statistics.
Conversely, extending the sample to higher redshift would continue the decline in angular host galaxy size and would further limit the reliability of these measurements and classifications.

\section{Comparison of phase-corrected and original \mbox{Si\,{\sc ii}} velocities}
\label{app:phase-correction}

\begin{figure*}
\centering
\includegraphics[width=0.85\textwidth]{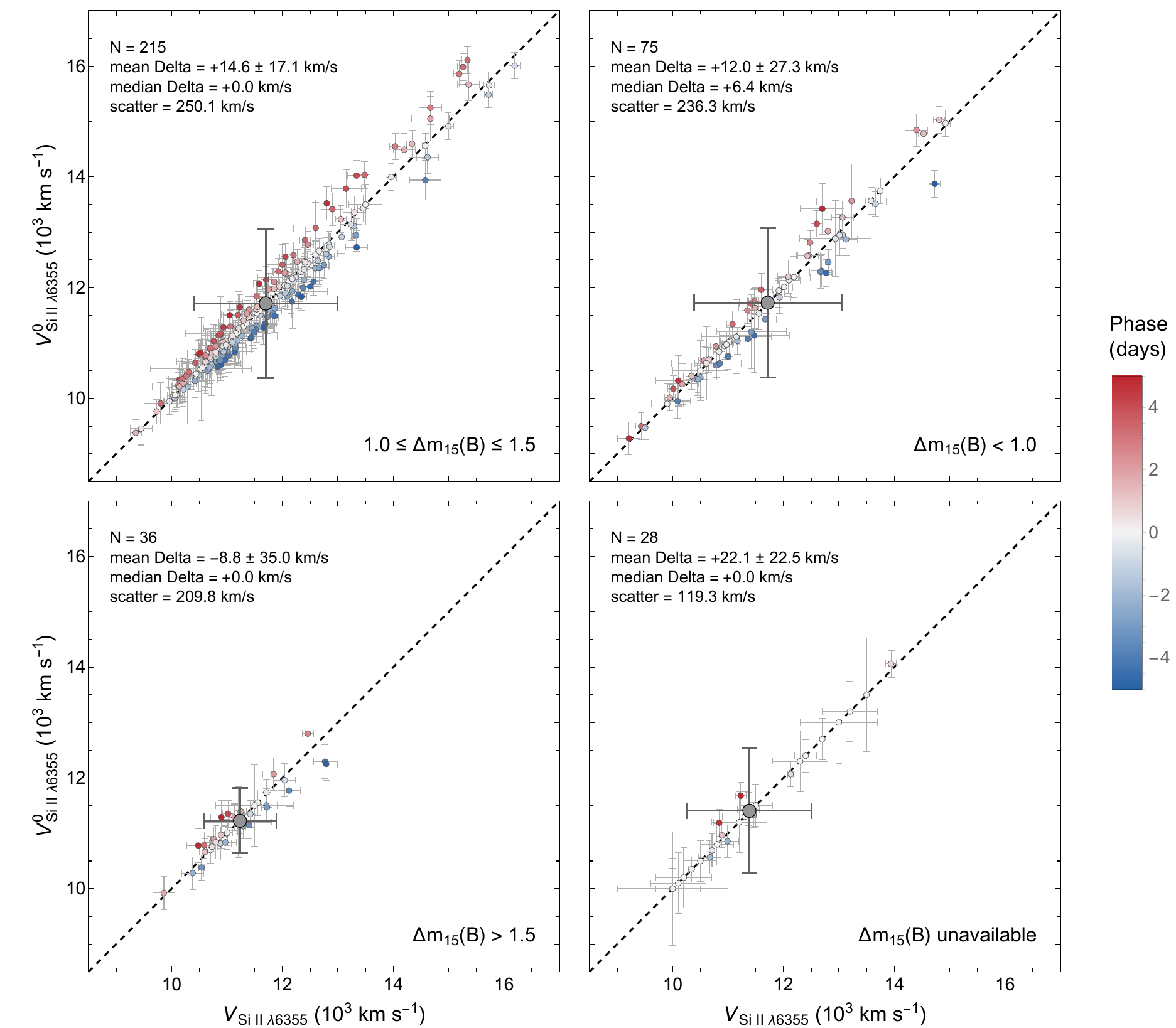}
\caption{Object-by-object comparison between the original \mbox{Si\,{\sc ii}} velocities, $V_{\rm Si}$, and the phase-corrected velocities at maximum light, $V_{\rm Si}^{0}$, for the 354 SNe~Ia in our sample. The panels show, respectively, SNe~Ia with $1.0\leq\Delta m_{15}(B)\leq1.5$~mag, $\Delta m_{15}(B)<1.0$~mag, $\Delta m_{15}(B)>1.5$~mag, and unavailable $\Delta m_{15}(B)$ values. The dashed line marks equality. Colours indicate the measurement phase relative to maximum light. Horizontal error bars show the reported uncertainties of the original velocity measurements; vertical error bars include the phase-correction scatter in quadrature. The large grey circles show the mean values in each panel, and their horizontal and vertical error bars represent the corresponding sample standard deviations.}
\label{fig:phase-correction}
\end{figure*}

To quantify the practical effect of the phase correction adopted in the main analysis, Fig.~\ref{fig:phase-correction} compares the original \mbox{Si\,{\sc ii}} velocity measurements, $V_{\rm Si}$, with the corresponding velocities at maximum light, $V_{\rm Si}^{0}$, obtained using the relation of \citet{2011ApJ...742...89F}.
This relation was calibrated for spectroscopically normal SNe~Ia with $1.0\leq\Delta m_{15}(B)\leq1.5$~mag
and for phases $-6\leq t\leq +10$~d.
Here, $\Delta m_{15}(B)$ is the decline in the $B$-band magnitude during the first 15~d after maximum light and is a standard light-curve decline-rate parameter for SNe~Ia.
The phase domain of our sample, $-5\leq t\leq +5$~d, lies entirely within this calibrated interval and is substantially narrower.
When several velocity measurements were available for an SN~Ia, the measurement closest to maximum light was selected; consequently, the phase distribution is concentrated around $t=0$ (see the inset of Fig.~\ref{Fits}).
Thus, the correction is applied only close to maximum light, where the velocity evolution is mild.

We carried out an additional literature search and compiled from different sources
$\Delta m_{15}(B)$ values for 326 of the 354 SNe~Ia,
i.e. $\sim92$ per cent of the full sample.
The comparison is shown separately for SNe~Ia with $1.0\leq\Delta m_{15}(B)\leq1.5$~mag, $\Delta m_{15}(B)<1.0$~mag, $\Delta m_{15}(B)>1.5$~mag, and unavailable $\Delta m_{15}(B)$ values.
The first group contains 215 SNe~Ia, while the other groups contain 75, 36, and 28 SNe~Ia, respectively.
We define the phase-correction difference as $\Delta V_{\rm Si}=V_{\rm Si}^{0}-V_{\rm Si}$.

For the 215 SNe~Ia within the formal $\Delta m_{15}(B)$ range of the \citet{2011ApJ...742...89F} relation, the mean correction with its standard error and the median correction are $+14.6\pm17.1$ and $+0.0$ ${\rm km~s^{-1}}$, respectively, with a scatter of $250.1$ ${\rm km~s^{-1}}$. For the 75 SNe~Ia with $\Delta m_{15}(B)<1.0$~mag, the corresponding values are $+12.0\pm27.3$, $+6.4$, and $236.3$ ${\rm km~s^{-1}}$, respectively. For the 36 SNe~Ia with $\Delta m_{15}(B)>1.5$~mag, they are $-8.8\pm35.0$, $+0.0$, and $209.8$ ${\rm km~s^{-1}}$, respectively. Finally, for the 28 SNe~Ia without available $\Delta m_{15}(B)$ measurements, the mean correction with its standard error and the median correction are $+22.1\pm22.5$ and $+0.0$ ${\rm km~s^{-1}}$, with a scatter of $119.3$ ${\rm km~s^{-1}}$.

Thus, for the phase-restricted sample used in this work, the phase correction produces only modest changes in the \mbox{Si\,{\sc ii}} velocities, with no systematic difference apparent among the four $\Delta m_{15}(B)$ groups. This behaviour is expected because all measurements are obtained close to maximum light, where the \mbox{Si\,{\sc ii}} velocity evolution is mild. The uncertainty of the original velocity is shown along the horizontal direction in Fig.~\ref{fig:phase-correction}. The vertical uncertainty includes both the reported velocity uncertainty and the phase-correction scatter reported by \citet{2011ApJ...742...89F}, added in quadrature.

\section{Analysis using uncorrected \mbox{Si\,{\sc ii}} velocities}
\label{app:uncorrected}

\begin{table*}
\centering
\begin{minipage}{155mm}
\caption{Comparison of the normalised galactocentric distance distributions using uncorrected \mbox{Si\,{\sc ii}} velocities.}
\tabcolsep 3pt
\label{tab:uncorrected-radial}
\begin{tabular}{ccccccccccrrrr}
\hline
\multicolumn{1}{c}{Morph} & \multicolumn{4}{c}{--------------- Subsample~1 ---------------} & \multicolumn{1}{c}{vs} & \multicolumn{4}{c}{--------------- Subsample~2 ---------------} & \multicolumn{1}{c}{$P_{\rm KS}^{\rm MC}$} & \multicolumn{1}{c}{$P_{\rm AD}^{\rm MC}$} & \multicolumn{1}{c}{$\mathrm{Med}(P_{\rm KS})$} & \multicolumn{1}{c}{$\mathrm{Med}(P_{\rm AD})$}\\
 & \multicolumn{1}{c}{Un/Targeted} & \multicolumn{1}{c}{SN} & \multicolumn{1}{c}{$N_{\rm SN}$} & \multicolumn{1}{c}{$\langle R_{\rm SN}/R_{25}\rangle$} && \multicolumn{1}{c}{Un/Targeted} & \multicolumn{1}{c}{SN} & \multicolumn{1}{c}{$N_{\rm SN}$} & \multicolumn{1}{c}{$\langle R_{\rm SN}/R_{25}\rangle$} && &&\\
\hline
All & All & NV & 236 & 0.41$\pm$0.02 & vs & All & HV & 118 & 0.39$\pm$0.03 & 0.553 & 0.546 & $0.626^{+0.236}_{-0.243}$ & $0.607^{+0.218}_{-0.228}$\\
All & Untargeted & NV & 98 & 0.35$\pm$0.03 & vs & Untargeted & HV & 45 & 0.37$\pm$0.05 & 0.791 & 0.682 & $0.664^{+0.197}_{-0.239}$ & $0.601^{+0.182}_{-0.177}$\\
All & Targeted & NV & 138 & 0.46$\pm$0.02 & vs & Targeted & HV & 73 & 0.41$\pm$0.03 & 0.174 & 0.153 & $0.387^{+0.206}_{-0.157}$ & $0.212^{+0.177}_{-0.104}$\\
\\
Elliptical & All & NV & 31 & 0.37$\pm$0.07 & vs & All & HV & 13 & 0.41$\pm$0.13 & 0.822 & 0.515 & $0.779^{+0.155}_{-0.262}$ & $0.514^{+0.182}_{-0.154}$\\
\\
S0/a--Sdm$^*$ & All & NV & 150 & 0.43$\pm$0.02 & vs & All & HV & 78 & 0.42$\pm$0.03 & 0.824 & 0.757 & $0.793^{+0.155}_{-0.250}$ & $0.810^{+0.125}_{-0.199}$\\
S0/a--Sdm$^*$ & Untargeted & NV & 55 & 0.37$\pm$0.04 & vs & Untargeted & HV & 24 & 0.48$\pm$0.07 & 0.492 & 0.175 & $0.363^{+0.189}_{-0.144}$ & $0.184^{+0.122}_{-0.081}$\\
S0/a--Sdm$^*$ & Targeted & NV & 95 & 0.47$\pm$0.03 & vs & Targeted & HV & 54 & 0.40$\pm$0.03 & 0.166 & 0.135 & $0.286^{+0.198}_{-0.134}$ & $0.148^{+0.127}_{-0.073}$\\
\hline\end{tabular}
\parbox{\hsize}{\emph{Notes:} The explanation of the $P$-values is the same as in Table~\ref{RSNR25}. Mean values and their standard errors are listed. $^*$ denotes $i<70^{\circ}$.}
\end{minipage}
\end{table*}

\begin{table*}\centering\begin{minipage}{155mm}
\caption{Comparison of host galaxy sizes ($R_{25}$ in kpc) using uncorrected \mbox{Si\,{\sc ii}} velocities.}
\tabcolsep 4pt\label{tab:uncorrected-size}
\begin{tabular}{ccccccccccrrrr}\hline
\multicolumn{1}{c}{Morph} & \multicolumn{4}{c}{------------ Subsample~1 ------------} & \multicolumn{1}{c}{vs} & \multicolumn{4}{c}{------------ Subsample~2 ------------} & \multicolumn{1}{c}{$P_{\rm KS}^{\rm MC}$} & \multicolumn{1}{c}{$P_{\rm AD}^{\rm MC}$} & \multicolumn{1}{c}{$\mathrm{Med}(P_{\rm KS})$} & \multicolumn{1}{c}{$\mathrm{Med}(P_{\rm AD})$}\\
 & \multicolumn{1}{c}{Un/Targeted} & \multicolumn{1}{c}{SN} & \multicolumn{1}{c}{$N_{\rm SN}$} & \multicolumn{1}{c}{$\langle R_{25}\rangle$} && \multicolumn{1}{c}{Un/Targeted} & \multicolumn{1}{c}{SN} & \multicolumn{1}{c}{$N_{\rm SN}$} & \multicolumn{1}{c}{$\langle R_{25}\rangle$} && &&\\\hline
All & All & NV & 236 & 15.6$\pm$0.4 & vs & All & HV & 118 & 16.9$\pm$0.7 & 0.181 & 0.144 & $0.157^{+0.180}_{-0.096}$ & $0.154^{+0.170}_{-0.088}$\\
All & Untargeted & NV & 98 & 13.1$\pm$0.7 & vs & Untargeted & HV & 45 & 13.0$\pm$0.9 & 0.851 & 0.906 & $0.735^{+0.200}_{-0.293}$ & $0.822^{+0.131}_{-0.261}$\\
All & Targeted & NV & 138 & 17.4$\pm$0.6 & vs & Targeted & HV & 73 & 19.3$\pm$0.8 & \textbf{0.049} & \textbf{0.042} & $\textbf{0.030}^{+\textbf{0.056}}_{-\textbf{0.021}}$ & $\textbf{0.040}^{+\textbf{0.055}}_{-\textbf{0.024}}$\\
\\
Elliptical & All & NV & 31 & 16.5$\pm$1.2 & vs & All & HV & 13 & 17.8$\pm$2.3 & 0.264 & 0.222 & $0.315^{+0.258}_{-0.178}$ & $0.227^{+0.150}_{-0.107}$\\
\\
S0/a--Sdm & All & NV & 188 & 16.2$\pm$0.5 & vs & All & HV & 97 & 17.6$\pm$0.7 & 0.269 & 0.163 & $0.181^{+0.207}_{-0.113}$ & $0.156^{+0.176}_{-0.091}$\\
S0/a--Sdm & Untargeted & NV & 68 & 14.2$\pm$0.7 & vs & Untargeted & HV & 33 & 14.5$\pm$1.0 & 0.607 & 0.955 & $0.793^{+0.147}_{-0.272}$ & $0.878^{+0.090}_{-0.190}$\\
S0/a--Sdm & Targeted & NV & 120 & 17.4$\pm$0.6 & vs & Targeted & HV & 64 & 19.2$\pm$0.9 & 0.168 & 0.108 & $0.104^{+0.141}_{-0.066}$ & $0.092^{+0.110}_{-0.054}$\\
\hline\end{tabular}
\parbox{\hsize}{\emph{Notes:} The explanation of the $P$-values is the same as in Table~\ref{RSNR25}.}
\end{minipage}\end{table*}

\begin{table*}\centering\begin{minipage}{165mm}
\caption{Comparison of host galaxy stellar masses ($\log(M_{\ast}/{\rm M_{\odot}})$ in dex) using uncorrected \mbox{Si\,{\sc ii}} velocities.}
\tabcolsep 3.2pt\label{tab:uncorrected-mass}
\begin{tabular}{ccccccccccrrrr}\hline
\multicolumn{1}{c}{Morph} & \multicolumn{4}{c}{--------------- Subsample~1 ---------------} & \multicolumn{1}{c}{vs} & \multicolumn{4}{c}{--------------- Subsample~2 ---------------} & \multicolumn{1}{c}{$P_{\rm KS}^{\rm MC}$} & \multicolumn{1}{c}{$P_{\rm AD}^{\rm MC}$} & \multicolumn{1}{c}{$\mathrm{Med}(P_{\rm KS})$} & \multicolumn{1}{c}{$\mathrm{Med}(P_{\rm AD})$}\\
 & \multicolumn{1}{c}{Un/Targeted} & \multicolumn{1}{c}{SN} & \multicolumn{1}{c}{$N_{\rm SN}$} & \multicolumn{1}{c}{$\langle\log(M_{\ast}/{\rm M_{\odot}})\rangle$} && \multicolumn{1}{c}{Un/Targeted} & \multicolumn{1}{c}{SN} & \multicolumn{1}{c}{$N_{\rm SN}$} & \multicolumn{1}{c}{$\langle\log(M_{\ast}/{\rm M_{\odot}})\rangle$} && &&\\\hline
All & All & NV & 236 & 10.66$\pm$0.04 & vs & All & HV & 118 & 10.71$\pm$0.06 & 0.435 & 0.559 & $0.471^{+0.342}_{-0.289}$ & $0.488^{+0.335}_{-0.289}$\\
All & Untargeted & NV & 98 & 10.45$\pm$0.09 & vs & Untargeted & HV & 45 & 10.50$\pm$0.10 & 0.195 & 0.363 & $0.669^{+0.233}_{-0.320}$ & $0.666^{+0.224}_{-0.283}$\\
All & Targeted & NV & 138 & 10.80$\pm$0.04 & vs & Targeted & HV & 73 & 10.85$\pm$0.07 & \textbf{0.046} & 0.109 & $0.190^{+0.299}_{-0.138}$ & $0.172^{+0.269}_{-0.117}$\\
\\
Elliptical & All & NV & 31 & 11.02$\pm$0.09 & vs & All & HV & 13 & 10.95$\pm$0.18 & 0.848 & 0.765 & $0.643^{+0.259}_{-0.305}$ & $0.588^{+0.271}_{-0.282}$\\
\\
S0/a--Sdm & All & NV & 188 & 10.70$\pm$0.04 & vs & All & HV & 97 & 10.76$\pm$0.05 & 0.111 & 0.240 & $0.258^{+0.350}_{-0.186}$ & $0.277^{+0.355}_{-0.187}$\\
S0/a--Sdm & Untargeted & NV & 68 & 10.57$\pm$0.08 & vs & Untargeted & HV & 33 & 10.61$\pm$0.10 & 0.738 & 0.767 & $0.729^{+0.200}_{-0.315}$ & $0.759^{+0.172}_{-0.268}$\\
S0/a--Sdm & Targeted & NV & 120 & 10.78$\pm$0.04 & vs & Targeted & HV & 64 & 10.84$\pm$0.06 & \textbf{0.043} & 0.108 & $0.184^{+0.287}_{-0.137}$ & $0.178^{+0.280}_{-0.122}$\\
\hline\end{tabular}
\parbox{\hsize}{\emph{Notes:} The explanation of the $P$-values is the same as in Table~\ref{RSNR25}.}
\end{minipage}\end{table*}

As an independent check of the phase-correction procedure, we repeated the velocity-dependent analyses using the original published \mbox{Si\,{\sc ii}} velocity measurements and their uncertainties in the phase range $-5\leq t\leq +5$ d, without applying the \citet{2011ApJ...742...89F} correction. The resulting sample contains 236 NV and 118 HV SNe~Ia. In each of $10^4$ MC realisations, both the original velocity and the relevant environmental parameter were perturbed according to their reported uncertainties, and the NV/HV classification was reassigned. To focus on the effect of the velocity definition, the tables below include only comparisons that depend on the NV/HV division; comparisons between discovery strategies and with external samples are omitted.

\subsection{Distribution of uncorrected \mbox{Si\,{\sc ii}} velocities}

We also repeated the bimodal Gaussian MLE analysis using the uncorrected velocities. For the full sample, the median NV and HV component weights are $72^{+4}_{-5}$ and $28^{+5}_{-4}$ per cent, the component means are $11100$ and $12900$ ${\rm km~s^{-1}}$, and the dispersions are $800$ and $1400$ ${\rm km~s^{-1}}$, respectively. For the targeted subsample, the corresponding component weights are $62^{+8}_{-11}$ and $38^{+11}_{-8}$ per cent. Thus, the two-component description and its principal parameters remain stable when the original measurements and their uncertainties are used.

\subsection{Galactocentric distances, host galaxy size, and stellar mass}

The main results presented in Sects.~\ref{RESults2} and \ref{RESults3}, i.e., the environmental comparisons of NV and HV SNe~Ia, remain qualitatively unchanged when the original uncorrected velocities are used instead of those corrected following \citet{2011ApJ...742...89F} (see Tables~\ref{tab:uncorrected-radial}--\ref{tab:uncorrected-mass}).

\section{Object-by-object comparison with literature measurements}
\label{app:paired}

\begin{figure}
\centering
\includegraphics[width=0.36\textwidth]{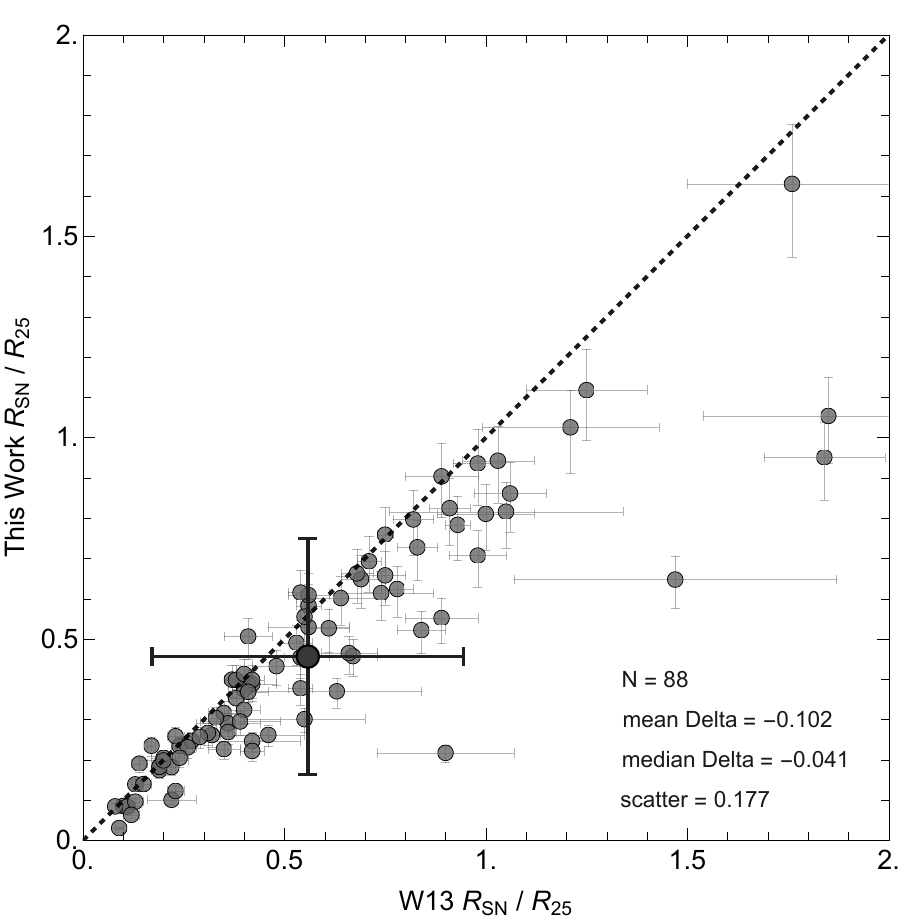}\\
\includegraphics[width=0.36\textwidth]{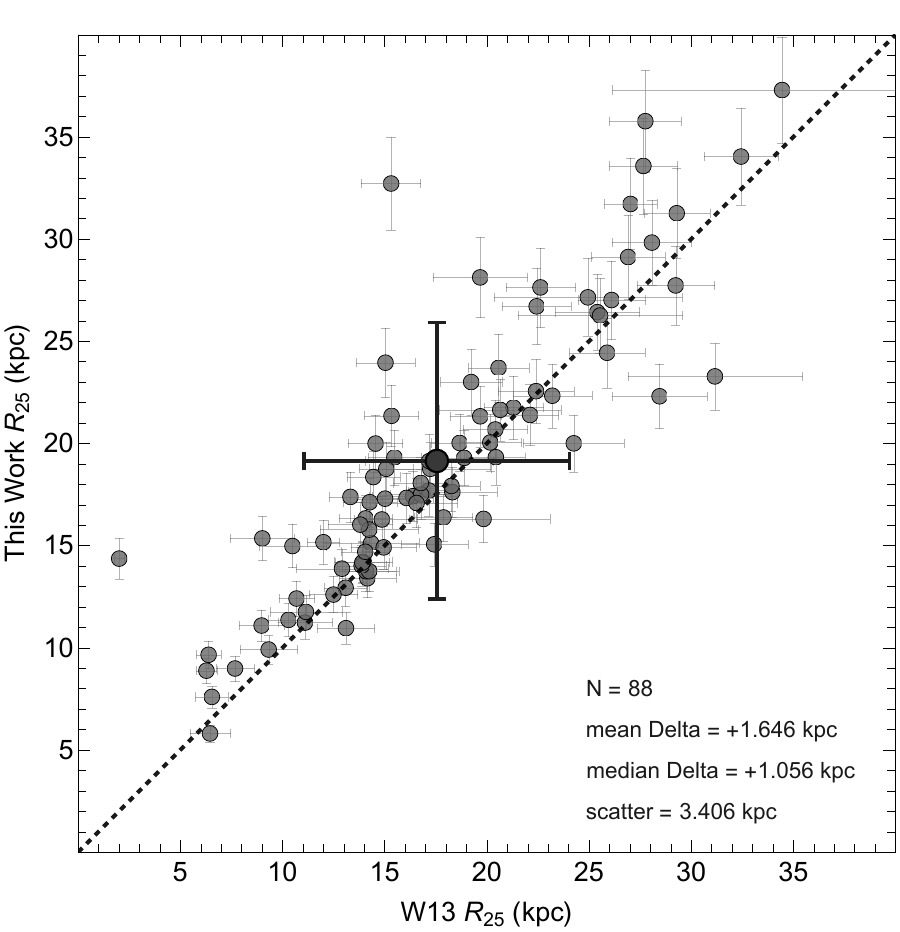}\\
\includegraphics[width=0.37\textwidth]{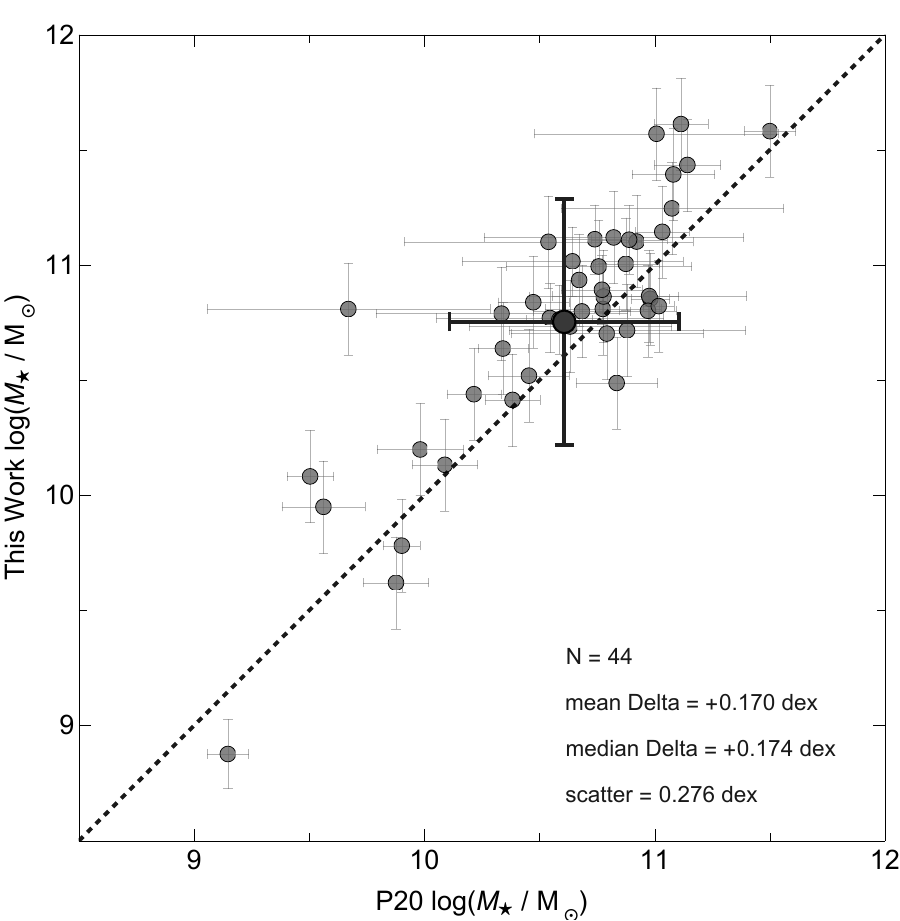}
\caption{Object-by-object comparison of the normalised galactocentric distances (upper panel)
and physical host sizes (middle panel) measured in this work and reported by \citetalias{2013Sci...340..170W} for
the 88 SNe~Ia common to both samples, as well as the comparison of the REGALADE and \citetalias{2020ApJ...895L...5P}
stellar masses for the 44 common host galaxies (bottom panel).
The dashed line represents equality, and the error bars show the reported uncertainties. The larger filled circle represents the mean values of the two compared distributions; its horizontal and vertical error bars indicate their corresponding sample standard deviations.}
\label{fig:AppCW13}
\end{figure}

To assess the agreement between individual measurements, we compare the normalised galactocentric distances and physical sizes for the 88 SNe~Ia common to \citetalias{2013Sci...340..170W}, and the stellar masses for the 44 hosts common to \citetalias{2020ApJ...895L...5P}.
In Fig.~\ref{fig:AppCW13}, the dashed lines show the one-to-one relation and the error bars represent the reported measurement uncertainties. For the normalised galactocentric distances, defining $\Delta(R_{\rm SN}/R_{25})=(R_{\rm SN}/R_{25})_{\rm this\ work}-(R_{\rm SN}/R_{25})_{\rm W13}$, the mean offset with its standard error and median offset are $-0.102\pm0.019$ and $-0.041$, respectively, with a scatter of $0.177$. For the physical sizes, defining $\Delta R_{25}=R_{25,\rm this\ work}-R_{25,\rm W13}$, the mean offset with its standard error and median offset are $+1.646\pm0.363$ and $+1.056$ kpc, respectively, with a scatter of $3.406$ kpc. For the stellar masses, defining $\Delta\log(M_{\ast}/{\rm M_{\odot}})=\log(M_{\ast}/{\rm M_{\odot}})_{\rm REGALADE}-\log(M_{\ast}/{\rm M_{\odot}})_{\rm P20}$, the mean offset with its standard error and median offset are $+0.170\pm0.042$ and $+0.174$ dex, respectively, with a scatter of $0.276$ dex.
Here, the scatter denotes the sample standard deviation of the object-by-object residuals.

The host galaxy parameters compared with those of \citetalias{2013Sci...340..170W} were obtained
using different data sources.
\citetalias{2013Sci...340..170W} compiled host galaxy parameters, including galaxy sizes, morphological types,
and inclinations, primarily from NED and HyperLeda, whereas in this work the corresponding host galaxy properties
were determined homogeneously using the imaging data and procedures described in Sect.~\ref{Hostdata}.
From the imaging data, $R_{25}$ can be determined with an uncertainty of less than 10 per cent,
based on the accuracy with which the galaxy major axis is measured;
the typical $R_{25}$ uncertainty in our sample is $\sim 7$ per cent and propagates directly into the uncertainty in $R_{\rm SN}/R_{25}$.
The systematic offsets in $R_{\rm SN}/R_{25}$ and $R_{25}$ (in kpc) may therefore partly reflect differences
in the adopted data sources and measurement procedures.

The stellar mass measurements compared here are also derived using different photometric data sets and methodologies.
\citetalias{2020ApJ...895L...5P} used SDSS $ugriz$ model magnitudes fitted with Z-PEG templates assuming a Salpeter IMF, whereas REGALADE adopts a calibrated mass--light--colour relation based on optical, 2MASS, and Wide-field Infrared Survey Explorer (WISE) photometry and is tied to a Chabrier IMF-based mass scale. Differences in photometric treatment, wavelength coverage, IMF, stellar-population modelling, and calibration can therefore contribute to the object-by-object offsets.

These object-by-object comparisons quantify offsets that may arise from differences in the adopted data sources, measurement procedures, and calibrations. They are therefore treated as external comparisons of independently derived quantities, rather than as evidence that the corresponding measurements are interchangeable or as a basis for the physical conclusions.

\section{Stellar mass analysis using REGALADE measurements}
\label{app:regalade-mass}

\begin{table*}
  \centering
  \begin{minipage}{165mm}
  \caption{Same as Table~\ref{Masses}, but using only the 342 host galaxy stellar masses available from REGALADE.}
  \tabcolsep 3.2pt
  \label{MassesRegalade}
    \begin{tabular}{ccccccccccrrrr}
    \hline
  \multicolumn{1}{c}{Morph} & \multicolumn{4}{c}{--------------- Subsample~1 ---------------} & \multicolumn{1}{c}{vs} & \multicolumn{4}{c}{--------------- Subsample~2 ---------------} & \multicolumn{1}{c}{$P_{\rm KS}^{\rm MC}$} & \multicolumn{1}{c}{$P_{\rm AD}^{\rm MC}$} & \multicolumn{1}{c}{$\mathrm{Med}(P_{\rm KS})$} & \multicolumn{1}{c}{$\mathrm{Med}(P_{\rm AD})$}\\
  & \multicolumn{1}{c}{Un/Targeted} & \multicolumn{1}{c}{SN} & \multicolumn{1}{c}{$N_{\rm SN}$} & \multicolumn{1}{c}{$\langle \log(M_{\ast}/{\rm M_{\odot}}) \rangle$} && \multicolumn{1}{c}{Un/Targeted} & \multicolumn{1}{c}{SN} & \multicolumn{1}{c}{$N_{\rm SN}$} & \multicolumn{1}{c}{$\langle \log(M_{\ast}/{\rm M_{\odot}}) \rangle$} && && \\
  \hline
     All & All & NV & 229 & 10.69$\pm$0.04 & vs & All & HV & 113 & 10.71$\pm$0.06 & 0.790 & 0.917 & $0.772^{+0.166}_{-0.255}$ & $0.818^{+0.141}_{-0.176}$\\
     All & All & All & 44 & 10.76$\pm$0.08 & vs & All & \citetalias{2020ApJ...895L...5P} & 44 & 10.59$\pm$0.06 & 0.208 & 0.107 & $0.207^{+0.258}_{-0.079}$ & $0.122^{+0.085}_{-0.053}$\\
     All & Targeted & All & 206 & 10.81$\pm$0.04 & vs & Untargeted & All & 136 & 10.52$\pm$0.06 & \textbf{0.008} & $<\textbf{0.001}$ & $\textbf{0.006}^{+\textbf{0.004}}_{-\textbf{0.002}}$ & $<\textbf{0.001}$\\
     All & Untargeted & NV & 94 & 10.54$\pm$0.08 & vs & Untargeted & HV & 42 & 10.48$\pm$0.11 & 0.210 & 0.381 & $0.369^{+0.239}_{-0.182}$ & $0.461^{+0.134}_{-0.123}$\\
     All & Targeted & NV & 135 & 10.80$\pm$0.04 & vs & Targeted & HV & 71 & 10.85$\pm$0.07 & 0.128 & 0.267 & $0.154^{+0.146}_{-0.081}$ & $0.274^{+0.134}_{-0.097}$\\
     \\
     Elliptical & All & NV & 31 & 11.06$\pm$0.09 & vs & All & HV & 12 & 10.82$\pm$0.18 & 0.306 & 0.274 & $0.311^{+0.207}_{-0.175}$ & $0.294^{+0.224}_{-0.169}$\\
     Elliptical & Targeted & All & 18 & 11.25$\pm$0.07 & vs & Untargeted & All & 25 & 10.81$\pm$0.12 & \textbf{0.003} & \textbf{0.002} & $\textbf{0.005}^{+\textbf{0.008}}_{-\textbf{0.003}}$ & $\textbf{0.003}^{+\textbf{0.003}}_{-\textbf{0.001}}$\\
     \\
     S0/a--Sdm & All & NV & 182 & 10.71$\pm$0.04 & vs & All & HV & 94 & 10.78$\pm$0.05 & 0.107 & 0.200 & $0.171^{+0.155}_{-0.092}$ & $0.218^{+0.127}_{-0.086}$\\
     S0/a--Sdm & Targeted & All & 180 & 10.80$\pm$0.03 & vs & Untargeted & All & 96 & 10.62$\pm$0.06 & 0.147 & \textbf{0.025} & $0.118^{+0.051}_{-0.041}$ & $\textbf{0.026}^{+\textbf{0.010}}_{-\textbf{0.008}}$\\
     S0/a--Sdm & Untargeted & NV & 66 & 10.62$\pm$0.07 & vs & Untargeted & HV & 30 & 10.61$\pm$0.11 & 0.860 & 0.874 & $0.884^{+0.093}_{-0.220}$ & $0.838^{+0.101}_{-0.116}$\\
     S0/a--Sdm & Targeted & NV & 116 & 10.76$\pm$0.04 & vs & Targeted & HV & 64 & 10.86$\pm$0.06 & 0.055 & 0.099 & $0.077^{+0.085}_{-0.045}$ & $0.130^{+0.080}_{-0.052}$\\
  \hline
  \end{tabular}
  \parbox{\hsize}{\emph{Notes:} The explanation of the $P$-values is the same as in Table~\ref{RSNR25}.}
\end{minipage}
\end{table*}

To assess whether the 12 stellar masses derived separately affect the conclusions of our stellar mass analysis, we repeated the comparisons in Table~\ref{Masses} using only the 342 SN~Ia hosts with stellar mass estimates available from REGALADE (Table~\ref{MassesRegalade}). In each of $10^4$ MC realisations, the individual \mbox{Si\,{\sc ii}} velocities and host galaxy stellar masses were perturbed according to their reported uncertainties. REGALADE adopts $H_0=71$ ${\rm km~s^{-1}~Mpc^{-1}}$, $\Omega_{\rm M}=0.27$, and $\Omega_{\Lambda}=0.73$, whereas \citetalias{2020ApJ...895L...5P} adopts $H_0=70$ ${\rm km~s^{-1}~Mpc^{-1}}$ and $\Omega_{\rm M}=0.30$. Over the redshift range of our sample ($z\leq0.04$), this cosmological difference corresponds to an offset of approximately $0.012$ dex in a luminosity-based stellar mass estimate. The 12 stellar masses derived in this work use $H_0=73$ ${\rm km~s^{-1}~Mpc^{-1}}$; relative to the REGALADE value, this corresponds to an offset of approximately $0.024$ dex. None of these 12 hosts is included in the 44-host comparison with \citetalias{2020ApJ...895L...5P}. Both offsets are much smaller than the typical REGALADE stellar mass uncertainties of $0.15$ dex for passive galaxies and $0.20$ dex for star-forming galaxies.

Overall, restricting the analysis to the 342 REGALADE stellar mass measurements preserves the principal NV/HV results, but affects the targeted--untargeted comparison of S0/a--Sdm hosts. Relative to Table~\ref{Masses}, the rejection status of the median KS and AD probabilities is unchanged for all other comparisons. For the targeted--untargeted S0/a--Sdm comparison, the median KS probability increases from $0.034$ in Table~\ref{Masses} to $0.118$ in Table~\ref{MassesRegalade}, and therefore no longer falls below the adopted significance threshold.
The corresponding median AD probability remains below this threshold ($0.005$ and $0.026$, respectively).
Thus, the targeted--untargeted difference for S0/a--Sdm hosts remains indicated by the AD test, while its KS-test significance is sensitive to the exclusion of the 12 separately derived stellar masses and may also reflect the reduced sample size.


\bsp	
\label{lastpage}
\end{document}